\documentclass[12pt]{article}

\usepackage{amsmath,amsfonts,amssymb,color,graphicx,overpic,}
\usepackage{slashed}
\usepackage{epsfig}
\usepackage[utf8]{inputenc}
\usepackage{amsmath}
\usepackage{ wasysym }
\usepackage{graphicx}
\usepackage[english]{babel}
\usepackage{graphic x}
\usepackage{relsize}
\usepackage{cite}

\begin{document}

\begin{titlepage}
\rightline{July 2017}
\vskip 2cm
\centerline{\large \bf
Dissipative dark matter halos:
The steady state solution}

\vskip 1.7cm
\centerline{R. Foot\footnote{
E-mail address: rfoot@unimelb.edu.au}}

\vskip 0.6cm
\centerline{\it ARC Centre of Excellence for Particle Physics at the Terascale,}
\centerline{\it School of Physics, University of Melbourne,}
\centerline{\it Victoria 3010 Australia}
\vskip 3.0cm
\noindent
Dissipative dark matter, where dark 
matter particle properties closely resemble familiar
baryonic matter, is considered.
Mirror dark matter, which arises from an isomorphic hidden sector, is 
a specific and theoretically constrained scenario. 
Other possibilities include models with more generic 
hidden sectors that contain massless dark photons (unbroken $U(1)$ gauge interactions).
Such dark matter not only features dissipative cooling processes, but is also assumed to have 
nontrivial heating sourced by ordinary supernovae (facilitated by the 
kinetic mixing interaction).
The dynamics of dissipative dark matter 
halos around rotationally supported galaxies, influenced
by heating as well as 
cooling processes, can be modelled by 
fluid equations. For a sufficiently isolated galaxy with stable star formation rate,
the dissipative dark matter halos are expected to evolve to a steady state configuration 
which is in hydrostatic equilibrium and 
where heating and cooling rates locally balance.
Here, we take into account the major cooling and heating processes, 
and numerically solve for the
steady state solution under the assumptions of spherical 
symmetry, negligible dark magnetic
fields, and that supernova sourced energy is transported to the halo
via dark radiation.
For the parameters considered, and assumptions made, we were 
unable to find a physically realistic solution for 
the constrained case of mirror dark matter halos. 
Halo cooling generally exceeds heating at realistic halo mass densities. This problem 
can be rectified in more generic dissipative dark matter models, and 
we discuss a specific example in some detail.

\end{titlepage}

\section{Introduction}

A variety of observations, on both large and small scales, suggest the existence of nonbaryonic 
dark matter in the Universe.
Anisotropies of the cosmic microwave background radiation, large 
scale structure, gravitational lensing 
observations, etc., are all consistent with cold or warm dark matter 
candidates, with $\Omega_{dm}/\Omega_b \approx
5.4$ from the analysis of PLANCK data \cite{plank}.
On smaller scales, dark matter properties can be probed in several ways including:
the abundance and distribution of satellite galaxies, and via the 
structural properties of galactic dark matter halos. 

The structural properties of galactic dark matter halos can be constrained 
by rotation curve measurements
of rotationally supported spiral and irregular galaxies, e.g. \cite{rubin,PS91,PPS,things}.
These measurements indicate that the dark matter halos around small dwarf irregular
and low surface brightness
galaxies can  dominate the gravitational potential at 
all measured radii, e.g. \cite{154,lss,littlethings}.  
The observed ($\sim$) linear rise of the rotational velocity in the inner region 
of such galaxies provides 
evidence that baryons must influence the dark matter structural properties, despite being only a small
subcomponent of the mass. In addition, various empirical scaling relations, including the Tully-Fisher 
relations \cite{tf,btf}, radial acceleration relation
\cite{Lelli,MC}, and others e.g. \cite{DS,Lelli2}, 
all support the notion that baryons influence dark matter structural properties.

This article concerns a type of self interacting dark matter, with dissipative interactions,
that has the potential to address these (and other) small scale structural
issues. The protype dissipative dark matter particle physics model assumes the existence
of a hidden sector   
that contains a `dark proton' and a `dark electron', coupled 
together via a massless `dark photon' \cite{sunny1}.
Such dark matter is dissipative in the sense that it can cool via the emission of dark photons.
The most theoretically constrained example of dissipative dark matter arises if
the hidden sector is exactly isomorphic to 
the ordinary sector (see e.g.\cite{footreview} for a review and
detailed bibliography). 
This kind of dark matter has been called
{\it mirror dark matter}, 
since the existence of an isomorphic hidden sector restores
improper Lorentz symmetries, including
space-time parity, as full symmetries of 
the Lagrangian describing fundamental particle interactions \cite{flv}.

In dissipative dark matter models, the halo around rotationally supported galaxies 
takes the form of a dark plasma with long range
interactions resulting in collective behaviour. This type of  halo dark matter 
can be modelled as a fluid
governed by Euler's equations, and if significant dark magnetic fields are present, by MHD equations.
The parameter space of interest is one where dissipation plays an important role:
in the absence of significant heating the
dark matter halo would collapse to a disk on a timescale shorter 
than the current Hubble time.
However, at the present epoch, such galactic dark matter can form an extended halo provided that
significant heat source(s) exist. 
The only viable heating mechanism identified so far 
is from ordinary Type II supernovae (SN), which can provide
heat sources for the dark sector if the kinetic mixing interaction exists \cite{sph,holdom,he}.

Dark matter galaxy halos are dynamical in this picture, influenced by the dissipative cooling as well
as supernova sourced heating. 
For a sufficiently isolated and unperturbed galaxy, the dark halo
is expected to evolve until it reaches an approximate steady state configuration, 
where the halo is in hydrostatic
equilibrium, and where heating and cooling rates locally balance.
This means that the current structural properties of 
the dark matter halos around such isolated galaxies
can be determined by the galaxy's  current baryonic properties, including
the SN abundance and distribution. 
Knowledge of the past history of the galaxy is not essential in this picture.

In this article  we aim to study the steady state solution for dissipative dark 
matter particle models,
including mirror dark matter, as well as  in more generic models. 
Previous work examined this problem, within the mirror dark matter context, 
considering only optically thin cooling, and neglected to
fully deal with 
capture and line emission \cite{foot13,foot14} (some work also looked at the optically thin case 
in more generic models \cite{sunny1,foot15,foot16,olga}).
Here, we aim to include all of the important cooling processes, and take into
account halo reabsorption of cooling radiation.
There are
some remaining simplifying assumptions: spherical symmetry and negligible 
dark magnetic fields. 
We have also followed earlier work \cite{foot13,foot14} and
assumed that the SN sourced energy is transported to the halo
via dark radiation.  The alternative case, where the SN sourced energy 
is transmitted to the halo via local collisional processes in the SN vicinity,
will be discussed in a separate paper.

For the parameters studied, 
we were unable to find physically realistic solutions 
for the constrained case of mirror dark matter halos. 
Halo cooling generally exceeds heating at realistic halo mass densities. 
This result prompts re-examination of the assumptions made; it also provides
motivation to explore more
generic dissipative models.
One such generic model, which features steady state solutions
with realistic halo mass densities, is studied in some detail.

This article is structured as follows. In section II we provide some background information
on dissipative dark matter models, focusing mainly on aspects of these models relevant
for understanding galaxy halo properties at the present epoch 
(the steady state solution of Euler's equations).  
In section III we discuss relevant halo properties: the ionization state,  
local heating and cooling rates.
In section IV we describe our method of solution of the steady state equations, give 
our results for the 
dissipative particle physics models studied, and
discuss. In section V we conclude. 


\section{Dissipative dark matter}

The standard model provides a remarkable description of the known elementary
particles and their fundamental interactions. There are no suitable
dark matter candidates in the standard model, but
this can easily be rectified by introducing a hidden sector. 
That is, a sector of additional particles and forces which couples to ordinary matter 
predominantly via gravity: 
\begin{eqnarray}
{\cal L} = {\cal L}_{SM} + {\cal L}_{dark} + {\cal L}_{mix}
\ .
\end{eqnarray}
If the hidden sector features an unbroken $U(1)$
gauge symmetry, then there will be `dark electromagnetic' interactions among the dark sector
particles  
mediated by a massless `dark photon'.
This kind of dark matter can  be dissipative in the sense that 
galactic dark matter halos can cool (in the absence of heating) on a timescale 
shorter than the current Hubble time.

\subsection{Two-component dissipative dark matter}

The protype dissipative dark matter model consists of a dark sector with two $U(1)'$ 
charged hidden sector particles:
a `dark electron' ($e_d$), and a `dark proton' ($p_d$). 
The interactions of these particles are governed by the Lagrangian,
\begin{eqnarray}
{\cal L}_{dark}  =  - \frac{1}{4}F^{'\mu \nu} F _{\mu \nu}^{'} + \bar{e}_{d}(iD _{\mu}\gamma ^{\mu} -
m_{e_d}) e_d 
+ \bar{p}_{d}(iD _{\mu}\gamma ^{\mu} - m _{p_d})p_d 
+ {\cal L}_{\text{mix}}
\ . 
\label{u7x}
\end{eqnarray}
Here, $F_{\mu \nu}^{'} \equiv \partial _{\mu} A _{\nu} ^{'} - \partial_{\nu} A _{\mu} ^{'}$ 
is the field strength tensor
associated with the dark $U(1)^{'}$ gauge interaction, and $A _{\mu}^{'}$ is the relevant gauge
field. The dark electron and dark proton are
described by the quantum fields $e_d, \ p_d$, and the covariant derivative is: 
$D_{\mu}  \equiv \partial _{\mu} + ig^{'}Q^{'} A_{\mu}^{'}$ (where $g^{'}$ is 
the coupling constant associated with this gauge interaction).
The Lagrangian [Eq.(\ref{u7x})] possesses
dark lepton and dark baryon global $U(1)$ symmetries which ensure that
the dark proton and dark electron are absolutely stable; this is
analogous to the way global baryon and lepton symmetries ensure the stability
of the electron and proton in the standard model.

The dark electron and dark proton are assumed to have $U(1)'$ charges opposite in sign, 
but their charge ratio $|Q'(p_d)/Q'(e_d)|$ is not necessarily unity; 
it is a fundamental parameter of the theory.
The self-interactions of the dark electron can be defined in terms of the dark fine structure constant, $\alpha_d
\equiv [g'Q'(e_d)]^2/4\pi$. 
A fundamental particle-antiparticle asymmetry is presumed to set the
relic abundance of dark electrons and dark protons
(that is, 
the relic abundance of dark antielectrons and dark antiprotons is negligible).

In addition to gravity, the dark sector interacts with the 
standard model particles via the kinetic mixing 
interaction involving the dark photon and the standard model hypercharge gauge boson \cite{he,flv}:
\begin{eqnarray}
{\cal L}_{\text{mix}} = \frac{\epsilon'}{2} \ F^{\mu \nu} F'_{\mu \nu}
\ .
\label{kine}
\end{eqnarray}
Here, $F^{\mu \nu}$ is the standard $U(1)_Y$ field strength tensor.
This renormalizable gauge-invariant interaction, characterized by 
the dimensionless parameter $\epsilon'$,
leads to photon - dark photon kinetic mixing, which 
imbues the dark electron and dark proton with an ordinary electric charge:
$- \epsilon e$ and $Z' \epsilon e$, where $Z' \equiv |Q'(p_d)/Q'(e_d)|$ \cite{holdom}. 
(Note that the parameter $\epsilon$, which is proportional to $\epsilon'$, 
is conveniently taken as the fundamental parameter.)
The new particle physics is fully described by the five fundamental 
parameters: $m_{e_d}, m_{p_d}, Z', \alpha_d$, and
$\epsilon$.

\subsection{Mirror model}

The standard model of particle physics has been very successful in describing the interactions
of the known elementary particles. 
Indeed,
the recent discovery of a Higgs-like scalar at the LHC \cite{atlas,cms}
is the latest chapter in this remarkable story.
An intriguing feature of 
the standard model
is that the
weak interaction
violates improper Lorentz symmetries, including parity and time reversal.
However, if the standard model is extended to include an exact `mirror' copy, that is, 
a duplicate set
of matter particles and gauge bosons, labelled henceforth with a prime ($'$), then
the improper Lorentz symmetries can be respected \cite{flv}.

In terms of a fundamental Lagrangian, the standard model is extended with an exact copy:
\begin{eqnarray}
{\cal L} = {\cal L}_{SM}(e,u,d,\gamma,...) + {\cal L}_{SM} (e', u', d', \gamma',...)
\ .
\end{eqnarray}
So far, no new fundamental parameters have been introduced. The elementary `mirror  particles'
have the same masses as their corresponding ordinary matter counterparts,  and their gauge self interactions
have the same coupling strength as the ordinary matter gauge self interactions. 
Since the mirror particles are described by a Lagrangian that is exactly 
analogous to that of the standard model,
there will be an entire set of `mirror elements': $H', He', Li', Be', B', C',...$
etc.,  the  properties of which will, of course, be completely analogous to the corresponding ordinary
elements: $H, He, Li, Be, B, C,...$.

The mirror sector particles are largely decoupled from the standard model particles, 
sharing only gravity, and 
possibly, additional interactions. Any additional interactions must respect the
gauge and space-time symmetries 
(including the improper space-time symmetries), conditions that lead to only two possible
renormalizable interactions. These are the Higgs - mirror 
Higgs portal interaction ($\lambda_h \phi^\dagger
\phi \phi'^{\dagger}\phi'$), and the kinetic mixing interaction of the form Eq.(\ref{kine}), where
$F^{\mu \nu}$ and $F'_{\mu \nu}$ are the field strength tensors of the standard model $U(1)_Y$ and 
mirror $U(1)'_Y$ gauge fields. 
The effect of kinetic mixing is to embellish the mirror sector  
particles with a tiny ordinary electric charge: $Q = -\epsilon e$ for the mirror electron, $e'$,
and $Q = \epsilon e$ for mirror proton, $p'$.

\subsection{Dark matter as mirror matter}

Mirror particles can be identified with the  nonbaryonic dark matter in  
the Universe, e.g. \cite{blin,hodges,foot04}.
On large scales, mirror dark matter closely resembles collisionless 
cold dark matter, e.g. successfully
reproducing the cosmic microwave background (CMB) anisotropy spectrum \cite{ber1,IV,ber2,footcmb}.
On smaller scales, though, the effects of the self 
interactions, and interactions with baryons via the kinetic mixing
interaction, lead to very different phenomenology.
Here, we provide a short overview of some relevant aspects of mirror dark matter.
A more comprehensive review, including a 
more detailed bibliography, can be found in  \cite{footreview}.

Early Universe cosmology of kinetically mixed mirror 
dark matter has been studied in a number of papers \cite{cg,paolo1,footc}.
In the early Universe, during the radiation dominated epoch, the ordinary and mirror particles 
form two almost decoupled sectors, each described by distinct temperatures, $T$ and $T'$.
Successful big bang nucleosynthesis (BBN) limits the 
energy density of the dark sector so that a temperature asymmetry
is required. An asymmetry is also needed to reproduce the CMB.\footnote{The origin
of the temperature asymmetry between the ordinary and mirror particles is unknown, 
but may potentially arise in chaotic inflation
models \cite{kolb,hodges,mohap}.} 
In fact, in the limit where $T' \ll T$,
mirror dark matter behaves like collisionless cold dark matter as far as BBN and CMB are concerned. 
It is well known that
collisionless cold dark matter can fit the measured CMB anisotropy spectrum,
with
$\Omega_{dm} \simeq 5.4\Omega_{b}$ 
obtained from the  analysis
of PLANCK data \cite{plank}
($\Omega$ is the usual normalized cosmic energy density parameter).
These considerations motivate the 
effective initial conditions at the BBN epoch:\footnote{
We assume that mirror dark matter comprises all of the dark matter in the Universe.
Hybrid dark matter models, where
a subdominant dissipative component is mixed with
a dominant collisionless
component, have  
been discussed in e.g. \cite{ddd,dddxyz,ddd2}.
}
\begin{eqnarray}
T' \ll T, \ \Omega_{b'} \simeq 5.4 \Omega_b
\ .
\label{ic4}
\end{eqnarray}

In the presence of nonzero kinetic mixing, entropy can be transferred from 
the ordinary sector to the mirror
particles, a process driven mainly by the particle interaction: $\bar e e \to \bar e' e'$. 
This entropy transfer ceases to be important for temperatures below the kinematic
threshold, $T \lesssim m_e$, and 
$T'/T$ asymptotes to \cite{paolo1,footc}: 
\begin{eqnarray}
T'/T \simeq 0.31\sqrt{\epsilon/10^{-9}}
\ .
\end{eqnarray}
The nonzero value of $T'/T$ appears to be rather important for the evolution of small scale
structure. Prior to mirror hydrogen recombination (at $T' \sim 1$ eV),
the growth of mirror dark matter density perturbations 
is impacted by dark acoustic oscillations and dark photon
diffusion. These effects can only be important for 
density perturbations characterized by length scales less than the sound horizon
at that time.

The effect of this dark sector physics is to severely suppress power on small scales.
This is somewhat analogous to the situation with warm dark matter, although the physical origin of the 
suppression involves very different physics.
This suppression of power on small scales can provide a simple explanation \cite{sunny2} 
for the observed deficit of 
satellite galaxies \cite{def1,def2}, and  potentially also a similar (albeit more modest)  
deficit observed for small field galaxies \cite{def3,def4,def5}.
Matching the relevant scales leads to a rough estimate of the fundamental kinetic mixing parameter: 
$\epsilon \sim 1 - 4 \times 10^{-10}$ \cite{sunny2}.

In this picture, very small scale perturbations are exponentially suppressed. 
So much so, that the smallest observable 
galaxies could only
have formed `top-down', that is, they arose out of the collapse of larger density perturbations.
If one contemplates the evolution of a galaxy mass scale perturbation,
then collapse occurs when the mean 
over-density of such a perturbation reaches a critical
value, $\delta \sim 1$.
During the nonlinear collapse process,
the dissipative dark matter is envisaged to form a disk-like structure. 
The collapse is not expected to be completely uniform,
perturbations at the edge of the dark disk could potentially
break off, and seed the formation of small satellite galaxies. 
In such a formation scenario, the satellite galaxies would 
have a planar and co-rotating distribution,
consistent, perhaps, with the 
properties of the observed satellites of the Milky Way \cite{sat1} and Andromeda \cite{sat2}.
Meanwhile, the bulk of the dark disk of the host galaxy might conceivably take the form
of a diffuse gas of dark sector particles.
It is envisaged that this
dark disk gas component would eventually disrupt due to the heating from ordinary 
supernovae (facilitated by the kinetic mixing interaction, to be discussed),  and ultimately,
expand to form a roughly spherical dark plasma halo.

Dark matter, if dissipative, might arise from a more generic hidden sector as 
opposed to the rather theoretically constrained case of mirror
dark matter. The two-component dissipative model, reviewed in section 2.1, is one such scenario.
That specific model has been examined in some detail 
in \cite{sunny1}, see also \cite{sunny2},
where some constraints on the fundamental parameters were derived.
Importantly, the  picture sketched above readily generalizes to this more generic case, 
and thus it remains a prime candidate for
dark matter that is able to explain structure on large scales and, potentially, also on small scales.

\section{Galaxy structure}

The dark matter halo around rotationally supported galaxies is envisaged to be
a dissipative plasma.\footnote{
The dark matter halo around elliptical and dwarf spheroidal galaxies is expected
to have very different physical properties in this picture. See \cite{sunny1,footreview} for
relevant  discussions.
} 
The bulk properties of such a plasma can be modelled
as a fluid governed by Euler's equations of fluid dynamics (and MHD equations
if dark magnetic fields play an important role).
These fluid equations take the form:
\begin{gather}
\frac{\partial \rho}{\partial t} + \nabla \cdot (\rho \mathbf{v}) = 0
\nonumber \ , \\
\frac{\partial \mathbf{v}}{\partial t} + (\mathbf{v} \cdot
\nabla)\mathbf{v} = -\left ( \nabla \phi + \frac{\nabla P}{\rho} \right)
\nonumber \ , \\
\frac{\partial}{\partial t} \left [\rho \left ( \frac{\mathbf{v}^2}{2} +
{\cal E} \right ) \right ] + \nabla \cdot \left [\rho \left (
\frac{\mathbf{v}^2}{2} + \frac{P}{\rho} + {\cal E} \right ) \mathbf{v}
\right ] - \rho \mathbf{v} \cdot \nabla \phi = {\cal H} - {\cal C}
\ .
\label{euler}
\end{gather}
Here $P$, $\rho$, and $\mathbf{v}$, denote the pressure, mass density, and
velocity of the fluid, and $\phi$ is the gravitational potential. 
${\cal E}$ is the internal energy per unit mass
of the fluid, so that $\rho \left (\mathbf{v}^2/2 + {\cal E} \right )$
is the energy per unit volume. Finally, ${\cal H}$ and ${\cal C}$ 
are the local heating and cooling rates per unit volume. 

Significant
simplifications occur if the system evolves to a steady state configuration;
the time derivatives vanish, and assuming there is no steady state
velocity flow, the system reduces to just two equations:
\begin{eqnarray}
\bigtriangledown  P &=& - \rho \nabla \phi  \ ,
\nonumber \\
{\cal H} &=& {\cal C}
\ .
\label{SSX}
\end{eqnarray}
These equations need to be satisfied at every location in the halo.
We shall assume that
the steady state configuration is the current physical state 
of rotationally supported galaxies that are sufficiently isolated
and have stable star formation rates.
Naturally, it would be useful to solve the
full system of time-dependent fluid equations to examine the evolutionary history 
and thereby check the consistency of this picture. This
would surely
require many details about the  galaxy's  properties and history etc., but nevertheless could
be attempted.

The halo density and temperature profiles  for which the steady state conditions [Eqs.(\ref{SSX})]
are satisfied represent a steady state solution.  
If this solution is unique, then the current halo properties 
are dictated in a large measure by the baryonic properties as the halo heating
is sourced by ordinary Type II supernovae (to be discussed).
This makes the dissipative dynamics highly predictive. Moreover, the tight coupling between the
physical properties of the halo and the galactic baryon content can potentially address long standing
indications of such a connection, e.g. \cite{MC,Lelli,DS,Lelli2,stacy}.
Previous work in this direction \cite{foot13,foot14,foot15,foot16,sunny1,olga} offers 
some encouragement 
that such a picture might lead to successful  phenomenology.

In this paper we endeavor to solve  equations Eq.(\ref{SSX})
to find steady state solutions
for mirror dark matter, as well as for the 
more generic dissipative dark matter model of section 2.1. We 
include all the major cooling processes, and take into 
account halo reabsorption of cooling radiation as
the optically thin approximation is not always valid. 
In fact, the wavelength-dependent finite optical depth will be taken into 
consideration 
for all
dark radiation sources, i.e. heating as well as cooling.
It turns out that the equations are somewhat nontrivial, and we do make the
simplifying assumption of spherical symmetry. 
While  there is reason to suppose that an approximately
spherically symmetric halo would form at large distances \cite{sunny1}, 
departures from spherical 
symmetry are anticipated
in the inner regions of galaxies. 
Nevertheless, we expect (as will be discussed) that such departures from
spherical symmetry are not of critical importance.
Naturally,
a more sophisticated analysis,
without the spherical symmetry assumption, could be undertaken following
essentially
the same procedure as developed here.

The baryons contribute to the gravitational potential, and their distribution is certainly not
spherical. In spiral galaxies the stellar distribution can 
be modelled as an azimuthally symmetric
disk with surface density \cite{freeman} 
\begin{eqnarray}
\Sigma (r) = m \ \frac{e^{-r/r_D}}{2\pi r_D^2}
\end{eqnarray}
where $m$ is the mass of the disk and $r_D$ is the disk scale length.
To have a mathematically consistent description we instead adopt a spherically 
symmetric distribution
for the baryons, with density defined by:
$\int^r_0 \rho^{stars}_{baryons}(r') 4\pi r'^2 dr' = \int_0^r \Sigma(r') 2\pi r' dr'$,
i.e. 
\begin{eqnarray}
\rho_{baryons}^{stars} (r) = m_{*} \ \frac{e^{-r/r_D}}{4\pi r_D^2 r}  
\ .
\label{doc}
\end{eqnarray}
Here, $m_*$ is the stellar mass parameterized in terms of a stellar 
mass fraction: $m_* = f_s m_{baryons}$.
In addition to stars, there is also a baryonic gas component - which generally features
a more spatially extended distribution.
We model
the gas density, $\rho_{baryons}^{gas}(r)$,  with an exponential profile
of the form Eq.(\ref{doc}), but with
$r_D^{gas} = 3r_D$ and total mass $m_{gas} = (1-f_s)m_{baryons}$.

The first equation in Eqs.(\ref{SSX}),  the
hydrostatic equilibrium condition, relates the dark matter fluid density and temperature. If  
we assume that all particle species are
in local thermodynamic equilibrium at a common temperature $T$, then 
the fluid pressure is $P = \rho T/{\bar m}$,  where
$\bar m \equiv \sum n_i m_i/\sum n_i$ is the mean mass of the particle species. 
[Each of these quantities is, of course, location dependent.]
For a spherically symmetric system, the hydrostatic equilibrium condition reduces to:\footnote{
Natural units with $\hbar = c = k_B=1$ are used unless otherwise indicated.}
\begin{eqnarray}
\frac{\partial T}{\partial r} = - \frac{T}{\rho} \frac{\partial \rho}{\partial r}\ +
\ \frac{T}{\bar m} \frac{\partial \bar m}{\partial r} \ - \ 
\bar m \nabla \phi
\label{eqmx}
\end{eqnarray}
where the gravitational acceleration is:
\begin{eqnarray}
\nabla \phi = \frac{G_N}{r^2} \ \int_0^r [\rho(r') + \rho_{baryons}(r')] \ 4\pi r'^2 dr'
\ .
\label{sunroast}
\end{eqnarray}
Here, $G_N$ is Newton's constant. 
Notice that we have included only the 
dark matter fluid density,  $\rho$, and 
the stellar, gas baryon components.\footnote{In addition to 
the diffuse dissipative fluid component, 
there can also be clumped dark matter objects, `dark stars'. In the analysis of this paper,
such a component is presumed to be subdominant and is, for simplicity, neglected.}
For a given fluid density [$\rho(r)$], and composition [$\bar m(r)$], 
the hydrostatic equilibrium condition can be solved for the
temperature profile if a boundary condition is specified. In our numerical work,
we take $dT/dr = 0$ at $r = 20r_D$.
[The results in the physical region of interest, $r \lesssim 6r_D$, are 
quite insensitive to the boundary condition, and its location, so long
as the boundary  is sufficiently distant.]

To proceed further, we need to evaluate the heating and cooling rates, also required is the
ionization state of the halo. 
We first evaluate these equations for a mirror dark matter halo; the modifications necessary for
the generic dissipative dark matter model will be subsequently indicated.
Since the heating 
and cooling rates depend on the ionization state,
and vice-versa, an iterative method will then be needed 
to solve this system of equations; one such method will be discussed in section 4.

\subsection{Ionization state of the halo}

In the mirror dark matter scenario,
the halo is a multicomponent plasma comprising a set of elements with varying degree's of ionization.
To simplify the discussion,  we shall, on occasion, adopt the notation: `electron' 
for `mirror electron', `photon' for
`mirror photon', etc. Since the discussion of the mirror dark matter 
plasma exclusively comprises mirror
particles with exactly analogous particle properties to 
ordinary matter, no confusion need arise.
We shall label the mirror elements and their ionization state with the notation:
$A^{k}$, where $A=H', \ He', C', O', Ne'....$ and $k = 0, 1,...,Z(A)$ 
represents the number of bound electrons present.
[$Z(A)$ is the nuclear charge, i.e. $Z(H')= 1, \ Z(He') = 2, \ Z(C') = 6 $, etc.]

The ionization state in a local region of interest 
is determined by the balancing of electron capture against the ionization processes:
\begin{eqnarray}
A^k + e' & \to & A^{k+1} + \gamma' \ \ \ {\rm Capture \ (free-bound\ transition)}\nonumber \\
A^k + e' & \to & A^{k-1} + e' + e' \ \ \ {\rm Electron \ impact \ ionization} \nonumber \\
A^k + \gamma' & \to & A^{k-1} + e' \ \ \ {\rm Photoionization} 
\ .
\end{eqnarray}
The cross sections for these processes will 
be denoted by $\sigma[A^k]_{fb}$, $\sigma[A^k]_{I}$,
and $\sigma[A^k]_{PI}$ respectively.
We introduce the notation $f_{A^k}$ for the 
fraction of $A$ states with $k$ bound electrons present. That is,
$n_{A^k} = f_{A^k} n_A$, where $n_A$ is the number density of all $A$ states. 
All these quantities are, of course, location dependent. At any given location,  
the rate of change of $n_{A^0}$ is:
\begin{equation}
\begin{aligned}
\frac{dn_{A^0}}{dt} = - n_{A^0} 
\int \frac{dn_{e'}}{dE_e} \sigma[A^0]_{fb}  v_e dE_e  
+ n_{A^1} 
\left[ 
\int \frac{dn_{e'}}{dE_e} \sigma[A^1]_{I}  v_e dE_e  
+ \int \frac{dF}{dE_\gamma}  \sigma[A^1]_{PI} dE_\gamma \right] 
\ .
\\[4pt]
\end{aligned}
\end{equation}
Here, $dF/dE_\gamma$ is the mirror photon flux at the location of interest, 
$dn_{e'}/dE_e$
is the local mirror electron energy distribution,
and $v_e = \sqrt{2E_e/m_e}$ is the mirror electron velocity.
In the steady state limit, $dn_{A^0}/dt \to 0$, and one finds:
\begin{eqnarray}
f_{A^1} = \frac{ 
\int \frac{dn_{e'}}{dE_e} \sigma[A^0]_{fb}  v_e dE_e}
{\int \frac{dn_{e'}}{dE_e} \sigma[A^1]_{I}  v_e dE_e  
 + 
\int \frac{dF}{dE_\gamma} \sigma[A^1]_{PI} dE_\gamma }  \ f_{A^0} 
\ .
\end{eqnarray}
More generally, using $dn_{A^k}/dt = 0$, one can deduce:
\begin{eqnarray}
f_{A^{k+1}} = \frac{ 
\int \frac{dn_{e'}}{dE_e} \sigma[A^k]_{fb}  v_e dE_e}
{\int \frac{dn_{e'}}{dE_e} \sigma[A^{k+1}]_{I}  v_e dE_e  
 + 
\int \frac{dF}{dE_\gamma} \sigma[A^{k+1}]_{PI} dE_\gamma }  \ f_{A^k} 
\label{trump}
\ .
\end{eqnarray}
This equation, together with $\sum_k f_{A^k} = 1$, determine the ionization state at
a given location in terms of the mirror electron distribution, 
mirror photon flux, and the relevant cross sections.
We now discuss  each of these three quantities in turn.

The electron distribution will be assumed to be Maxwellian:
\begin{eqnarray}
\frac{dn_{e'}}{dE_e} = n_{e'} 
\frac{2}{T}\sqrt{\frac{E_e}{\pi T}} \ e^{-E_e/T}
\ .
\end{eqnarray}
This is an important simplification.  In general,
significant departures from a Maxwellian distribution can occur
in the low density plasma environment from a variety of (typically) complex processes. 
One such process arises due to the halo heating mechanism assumed.
As will be discussed in more detail in section 3.3,
the kinetic mixing interaction transforms ordinary supernovae
into powerful heat sources. Supernovae 
generate energetic mirror photons which are absorbed in the 
halo via the photoionization process.
The ejected mirror electron resulting from photoionization  
can be very energetic, and thermalizes primarily by scattering off free and 
bound mirror electrons in the plasma.
The ionization due to such non-thermal scattering off bound mirror electrons
is neglected in our analysis, but could be important at low halo temperatures where there
are relatively few free mirror electrons.

The mirror photon flux originates from several sources: supernovae, line emission, 
capture, and bremsstrahlung.
In a given volume element, $dV$, the
differential luminosity of photons that arises from all 
of these sources combined will be denoted as
$d{\cal L}^S/dVdE_\gamma$. 
To calculate the flux of photons at a particular location in the halo we must integrate 
this luminosity over all possible source
locations and take into account reabsorption processes.

\begin{figure}[t]
    \centering
    \includegraphics[width=0.33\linewidth,angle=0]{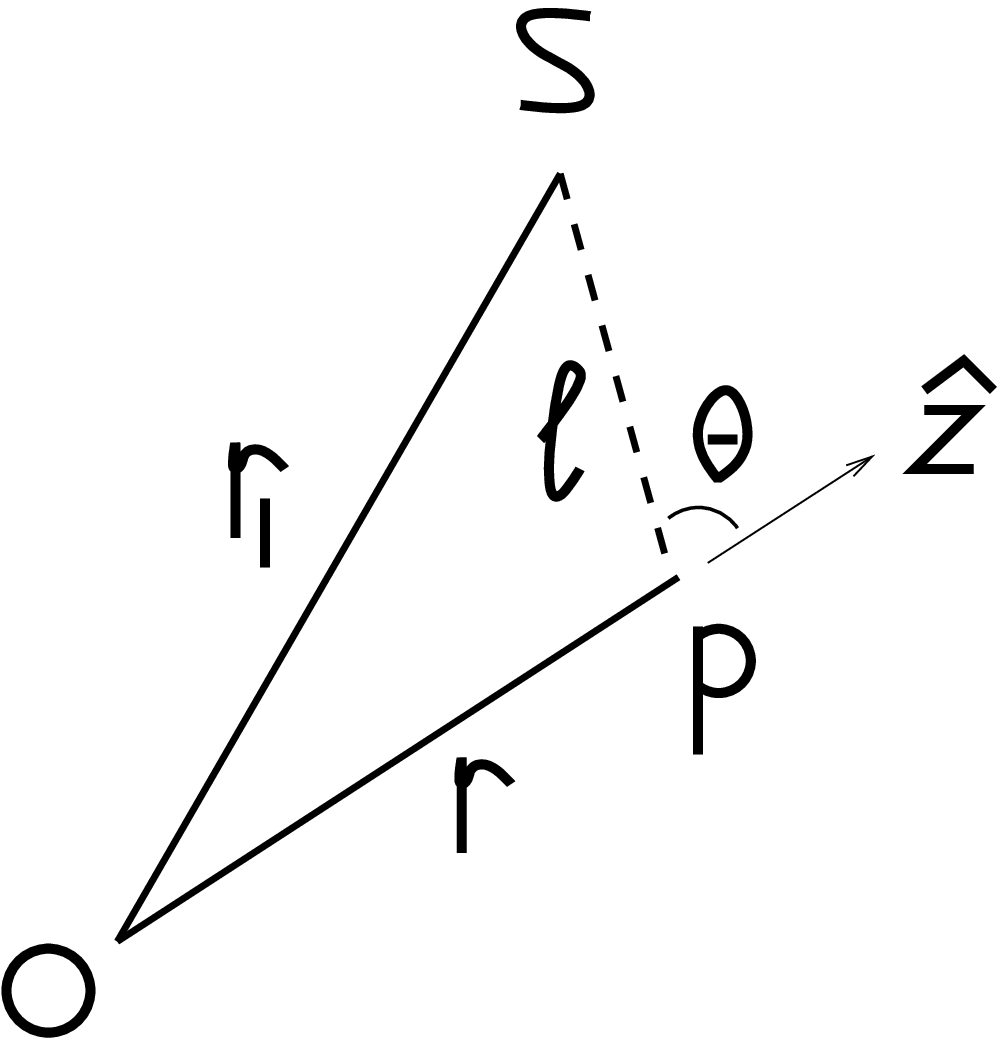}
\end{figure}

Within a spherical coordinate system with origin at the galaxy center, $O$, consider
a particular halo location of interest, $P$, at ${\bf r}$ in this coordinate system 
(see diagram). 
Some of the photons 
which arrive at $P$ propagate from a source location, $S$, at ${\bf r_1}$. 
It is convenient to define
a second spherical coordinate system (with coordinates $\ell, \theta, \phi$)
with origin now at $P$ (and with $z-axis$ in the direction of ${\bf r}$).
The photon flux at $P$ (i.e. at the origin of the second spherical coordinate system) 
can be found by integrating over all source locations. Taking into account 
the absorption along the photon path from $S$ to $P$, we have: 
\begin{eqnarray}
\frac{dF({\bf r})}{dE_\gamma}
 &=&  \int \int  \ \frac{1}{E_\gamma}\frac{d{\cal L}^S (r_1)}{dVdE_\gamma}  \  \frac{e^{-\tau} }{4\pi \ell^2} 
\ 2\pi \ell^2 d\cos\theta d\ell
\nonumber \\
 &=&  \int \int  \ \frac{e^{-\tau}}{2E_\gamma}\frac{d{\cal L}^S (r_1)}{dVdE_\gamma}  
\  d\cos\theta d\ell
\label{flux1}
\end{eqnarray}
where $r_1 = \sqrt{r^2 + \ell^2 + 2 r \ell \cos\theta}$ follows from this 
geometry, and we have set $r_1 = |{\bf r_1}|$, $r = |{\bf r}|$, etc.
The optical depth, $\tau = \tau(\ell,\theta,E_\gamma)$, is given by
\begin{eqnarray}
\tau (\ell,\theta,E_\gamma) = \sum_{A,k} \int_0^{\ell} \sigma[A^k]_{PI} \ n_{A^k}({\bf r},{\ell_1},\theta) \ 
d{\ell_1}
\label{17z}
\end{eqnarray}
where $n_{A^k}({\bf r},{\ell_1},\theta) \equiv n_{A^k}(r')$ with $r' =
\sqrt{r^2 + {\ell_1^2} + 2 r 
{\ell_1} \cos\theta}$.
The photoionization cross section in Eq.(\ref{17z}) is also a function of $r, \ell_1, \theta$, because it depends on
the temperature along the path: $T(r')$.
In deriving Eq.(\ref{flux1}) azimuthal symmetry has been used to perform the trivial
$\phi$ integration. [Azimuthal symmetry in the coordinate system
with origin at the point of interest, $P$, follows from spherical symmetry in the coordinate system
with origin at the galactic center.]

We now consider the cross sections. 
In addition to the $\sigma[A^k]_{fb}$, $\sigma[A^k]_I$, and $\sigma[A^k]_{PI}$ cross sections, 
we will also need the electron
excitation cross section, $\sigma[A^k]_{nln'l'}$, and the flux of bremsstrahlung photons.
For the electron excitation process, we made extensive use of 
the cross sections calculated by Group T-4  
of the Los Alamos National Laboratory, 
publicly available via their online web interface \cite{lanl}.
The LANL code is based on the method of Mann \cite{mann},
and calculated using
the first order many body theory.
The LANL code also makes use of the
Hartree-Fock method of R.D. Cowan \cite{Cowanbook}, 
developed at Group T-4 of 
the Los Alamos National Laboratory.

Our numerical work
used the LANL cross sections for the electron excitation process,
with a total of around $\sim 300$ of the most important $nl \to n'l'$ transitions
considered.
For the generic dark matter model, these cross sections will need to be  scaled to investigate
parameters $m_{e_d}, \alpha_d$ that are different from $m_e, \alpha$.
To understand the appropriate scaling with respect to these parameters an
analytic form for the electron excitation cross section is also useful. 
For hydrogen-like ions, Van Regemorter  \cite{VR} calculated  the cross section for 
electron impact excitation 
$nl \to n'l'$ (with excitation energy $E_{nln'l'}$) in the Bethe approximation:
\begin{eqnarray}
\sigma[A^k]_{nln'l'} = 
\pi a_0^2 \ \frac{8\pi f_{nln'l'}}{\sqrt{3}} \left[ \frac{Ry}{E_{nln'l'}}\right]^2 \frac{G(x)}{x}
\ \Theta(E_e - E_{nln'l'})
\end{eqnarray}
where $\Theta (y)$ is the Heaviside step function.
Also, $a_0 \equiv 1/(m_e \alpha)$, $f_{nln'l'}$ is the absorption oscillator strength, $Ry \equiv \alpha^2 m_e/2$, 
$x \equiv E_e/E_{nln'l'}$, and
$G(x)$ is the effective Gaunt factor of order unity. 

For electron impact ionization ($A^k + e' \to A^{k-1} + e'  + e'$) we have adopted
the  Lotz formula \cite{Lotz}:
\begin{eqnarray}
\sigma[A^k]_I = \sum_i  \ \frac{0.62 \alpha^2 \pi}{E_e I_i} \ ln \left( \frac{E_e}{I_i} \right) \ \Theta(E_e - I_i)
\end{eqnarray}
where $I_i$ denotes the ionization energies of 
the $k$ bound electrons ($i=1,...,k$) in the $A^k$ ion. The set of ionization energies, $I_i$, for each
$A^k$ ion  were acquired from the LANL web interface \cite{lanl}.

For photoionization, also called bound-free transition 
($A^k + \gamma' \to A^{k-1} + e'$), we used the Karzas and Latter result \cite{KL,RL} :
\begin{eqnarray}
\sigma[A^k]_{PI} = \sum_i \ \frac{32\pi \alpha I^2_i g_{bf}}{3\sqrt{3}m_e E_{\gamma}^3 n_i} \ \Theta(E_{\gamma} - I_i)
\ .
\end{eqnarray}
Here, the index $[i]$ represents the state of the electron prior to its ejection 
from the atom, $n_i$ is its principal quantum number, and $I_i$ is the ionization energy. 
The sum runs over all bound electrons [i.e. from $i=1,...,k$]. 
Near threshold, the Gaunt factor is unity to within 20\%, and we set
$g_{bf}=1$ in the numerical work.

For electron capture, also known  as free-bound transition ($A^k + e' \to A^{k+1} + \gamma'$), 
we used the modified Kramers formula
\cite{Pratt} :
\begin{eqnarray}
\sigma[A^k]_{fb} =    \sum_{i} \frac{16\pi \alpha I^2_i w_i g_{fb}}{3\sqrt{3} m_e^2  (E_e + I_i)  E_e n_i} 
\end{eqnarray}
where $w_i$ is the number of unoccupied states in the $n_i$ shell of the ion before recombination.
In the numerical work, we considered capture to the valence shell along with the next  
higher shell ($n_i=n_{valence}, n_i=n_{valence}+1$).
Again, we set the Gaunt factor $g_{fb}$ to unity.

The discussion above was relevant for mirror dark matter. 
The situation with  more generic 
dissipative models is quite analogous.
For the two-component model of section 2.1, 
each particle process: capture, excitation, ionization etc.,
has an analogue with the obvious replacements: $e' \to e_d$, $A^k \to p_d^k$,
$\gamma' \to \gamma_d$. The cross sections for these processes
are defined in terms of a new set of fundamental parameters:
$m_{e_d}, m_{p_d}, \alpha_d, Z', \epsilon$. We need to determine how the relevant cross sections
depend on these parameters.

Actually, if  $m_{e_d} \ll m_{p_d}$, then 
to a very good approximation the relevant cross sections depend only on $m_{e_d}, \alpha_d, Z'$.
For $Z'$ integer, we can choose the corresponding element $A$ with $Z' = Z(A)$ (e.g. for 
$Z'=6$ we take $A=C$).
With the element $A$ chosen, the mirror dark matter $A^k$ 
cross sections can be scaled to take into account
values of $m_{e_d}, \alpha_d$ different from $m_e, \alpha$. The result is the scaling:
\begin{eqnarray}
\left\{ \sigma[A^k]_{nln'l'}, \ E_e \right\} &\to &   
\left\{ \sigma[A^k]_{nln'l'}\frac{\alpha^2}{\alpha^2_d} \frac{m_e^2}{m^2_{e_d}},
\ E_e \frac{\alpha^2_d}{\alpha^2} \frac{m_{e_d}}{m_e} \right\} 
\nonumber \\
\left\{ \sigma[A^k]_{I}, \ E_e \right\} &\to   &
\left\{ \sigma[A^k]_I \frac{\alpha^2}{\alpha^2_d} \frac{m_e^2}{m^2_{e_d}},
\ E_e \frac{\alpha^2_d}{\alpha^2} \frac{m_{e_d}}{m_e} \right\} 
\nonumber \\
\left\{ \sigma[A^k]_{PI}, \ E_\gamma \right\} &\to   &
\left\{ \sigma[A^k]_{PI} \frac{\alpha}{\alpha_d} \frac{m_e^2}{m^2_{e_d}},
\ E_\gamma \frac{\alpha^2_d}{\alpha^2} \frac{m_{e_d}}{m_e} \right\} 
\nonumber \\
\left\{ \sigma[A^k]_{fb}, \ E_e \right\} &\to   &
\left\{ \sigma[A^k]_{fb} \frac{\alpha_d}{\alpha} \frac{m_e^2}{m^2_{e_d}},
\ E_e \frac{\alpha^2_d}{\alpha^2} \frac{m_{e_d}}{m_e} \right\} 
\ .
\end{eqnarray}
The
ionization energies also scale:
\begin{eqnarray}
I_i \to I_i \ \frac{\alpha^2_d}{\alpha^2} \frac{m_{e_d}}{m_e}
\ .
\end{eqnarray}
For the bremsstrahlung process, we only need to know how  the differential cooling rate
scales with $m_{e_d}, \alpha_d$, which can  
be gleaned from the explicit 
expression for this rate given in the following subsection.

\subsection{Cooling rates}

There are three sources of cooling that need to be taken into account: Line emission, capture, 
and bremsstrahlung.
In addition, conduction and convection processes can also contribute to the local 
cooling/heating rates.
These processes could be important if significant temperature gradients exist.
It turns out that the halo temperature profile derived from the steady
state conditions is close to isothermal, so that neglect of these 
processes might be justifiable. In any case, conduction/convection processes
will not be included in the analysis
presented here.

For thermal bremsstrahlung, also called free-free emission,
we follow the classical treatment of \cite{RL}.
The differential rate of energy radiated per unit volume due to electron scattering off ions of charge $Z_i$,
assuming a Maxwellian electron velocity distribution, is: 
\begin{eqnarray}
\frac{d{\cal C}_{ff}}{dE_\gamma} = \frac{16 \alpha^3}{3 m_e}
\left(\frac{2\pi}{3m_e T}\right)^{1/2} 
Z_i^2 n_i n_e e^{-E_\gamma/T} \ \bar g_{ff}
\ .
\end{eqnarray}
Here, $\bar g_{ff}$ is the velocity averaged Gaunt factor,
which can be approximated by the simple analytic expression \cite{RL}:
\begin{eqnarray}
\bar g_{ff} = \begin{cases}
\frac{\sqrt{3}}{\pi} \ ln \left[ \frac{4}{\xi} 
\frac{T}{E_\gamma}\right]
& \ \ {\rm for} \ E_\gamma < T \ , \\
\sqrt{\frac{3T}{\pi E_\gamma}} 
& \ \  {\rm for} \ E_\gamma > T\ .
\end{cases}
\end{eqnarray}
Here $\xi \simeq 1.781$ is Euler's constant.
This simple analytic form for the Gaunt factor is known to be valid
for $T \gtrsim Z^2 Ry$, where $Ry = 13.6$ eV.
For small dwarf galaxies where  
$T \lesssim Z^2 Ry$, the Gaunt factor is less accurate but still provides a reasonable estimate
for our purposes, especially as the bremsstrahlung cooling rate in small galaxies turns out to be 
much smaller than the other cooling processes.

The differential rate of energy radiated due to electron capture by an ion, $A^k$, is:
\begin{eqnarray}
\frac{d{\cal C}_{fb}}{dE_\gamma} = 
n_{A^k}\frac{dn_{e'}}{dE_e} v_e \sigma[A^k]_{fb} E_\gamma \ . 
\end{eqnarray}
Note that energy conservation implies $E_\gamma = E_e + I_i$.
Assuming that the electron's energy distribution is Maxwellian, we have:
\begin{eqnarray}
\frac{d{\cal C}_{fb}}{dE_\gamma} = 
2\sqrt{\frac{2}{m_e \pi}}\left(\frac{1}{T}\right)^{3/2} 
e^{-E_e/T} \ \sigma[A^k]_{fb} E_e E_\gamma
\ .
\end{eqnarray}

Energy is also radiated from line emission. Electrons can scatter off a 
bound electron in an ion, $A^k$, leading to the atomic transition $nl \to n'l'$ (with
excitation energy $E_{nln'l'}$).
The resulting energy radiated following de-excitation
is:
\begin{eqnarray}
\frac{d{\cal C}_{lines}}{dE_\gamma} = \sum n_{e'} n_{A^k}
\langle \sigma[A^k]_{nln'l'} 
 v_e \rangle \delta(E_\gamma - E_{nln'l'}) E_{nln'l'}
\end{eqnarray}
where the sum runs over the $nl$ quantum numbers that correspond to each of the $k$ bound electrons,
and all possible $n'l'$ quantum numbers of the atomic excitations (and 
also over all $A^k$ ions).
For a Maxwellian
electron velocity distribution, 
\begin{eqnarray}
\langle \sigma[A^k]_{nln'l'} v_e \rangle 
= 2\sqrt{\frac{2}{m_e \pi}}\left(\frac{1}{T}\right)^{3/2} 
\int_{E_{nln'l'}}^{\infty} \sigma[A^k]_{nln'l'} \ e^{-E_e/T} \ E_e dEe
\ .
\end{eqnarray}

The differential rate of radiation energy loss per unit volume
at a location $P$ (at position ${\bf r}$) in the halo is 
the sum of these three contributions:
\begin{eqnarray}
\frac{d{\cal C}({\bf r})}{dE_\gamma} = \frac{d{\cal C}_{ff}({\bf r})}{dE_\gamma}
+ \frac{d{\cal C}_{fb}({\bf r})}{dE_\gamma}
+ \frac{d{\cal C}_{lines}({\bf r})}{dE_\gamma}
\ .
\label{wed1}
\end{eqnarray}
These dark photons, together with those originating from 
Type II supernovae (to be considered in more detail in the following subsection),
contribute to the differential source flux, $d{\cal L}^S/dV dE_\gamma$, which influences the ionization state of
the halo [Eqs.(\ref{flux1},\ref{trump})]. Some of these cooling photons will be reabsorbed 
and also affect the heating rate, ${\cal H}$ (to be discussed
shortly).

The cooling rate will depend on the relative abundances of the various mirror elements.
Mirror BBN calculations
with the initial conditions of Eq.(\ref{ic4})
and with $\epsilon \sim 2\times 10^{-10}$
(as suggested by the observed deficit of satellite galaxies \cite{sunny2})
have concluded that the primordial mirror helium abundance dominates over mirror hydrogen, consistent with
general arguments \cite{ber1}, with
the helium mass fraction $Y'_p \approx 0.95$ \cite{paolo2} (see also Figure 3.4 of \cite{footreview}).
Heavier mirror elements are expected to 
be synthesized in mirror stars at an early epoch cf. \cite{berstar}.
Unfortunately, the detailed chemical composition of the mirror sector resulting from 
stellar evolution at the early epoch is rather
difficult to surmise: It involves
unfamiliar initial conditions, 
chemical composition etc., including unknown quantities such as initial mass function.
We shall assume for simplicity that the composition of the mirror metal component
is the same as the solar abundance of the corresponding ordinary elements, but allow
for an overall scale factor $\zeta$ for the metal component.  
Naturally, modifications of the relative abundances of the various
elements could be looked at it.
In Table 1 the standard solar abundances of the ordinary elements are given, along with
modified abundances incorporating the  
higher primordial mirror helium abundance ($Y'_P \approx 0.95$, 
suggested by mirror BBN calculations, translates to $log [n_{He}/n_{H}] \simeq 0.68$).
The parameter $\zeta$ allows adjustment of the mirror
metal fraction, and we consider a wide range in our numerical work ($-2.0 < \zeta < 2.0$).  

\begin{table}
\centering
\begin{tabular}{c c}
\hline\hline
Element & $log(n/n_H)$ (solar/modified)  \\
\hline
He & -1.01/0.68 \\
C & -3.44   + $\zeta$  \\
O & -3.07   + $\zeta$  \\
Ne & -3.91  + $\zeta$  \\
Si & -4.45  + $\zeta$  \\
Fe & -4.33  + $\zeta$  \\
\hline\hline
\end{tabular}
\caption{\small Solar abundances ($\zeta = 0$) from \cite{anders} and modified abundances.}
\end{table}

As a check of our code, we have computed the cooling function, 
$\Lambda_N \equiv  {\cal C}/(n_e n_t)$ [where $n_t \equiv \sum_A n_A$ is the total
number density of mirror ions], for the idealized case of a low density
optically thin plasma with ionization dominated by electron impact. 
In this circumstance the ionization state and
cooling function depend only on the local temperature.
Adopting solar abundances, but choosing $\zeta = 0.1$ to compensate for the restriction
of just five metal components, we found the 
cooling function shown in Figure 1. This cooling function compares 
reasonably well with more accurate results found 
in the literature, such as the result of Dopita and Sutherland \cite{ds9}. 
In the numerical work to follow we use the modified abundances, which
take into account the $log[n_{He'}/n_{H'}] \approx 0.68$
estimated from mirror BBN.
\begin{figure}[t]
\centering
\includegraphics[width=0.48\linewidth,angle=270]{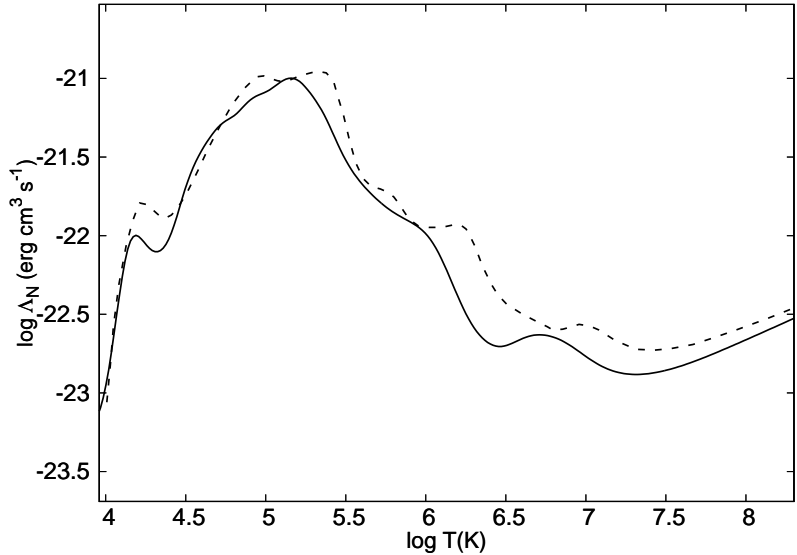}
\caption{
\small
Optically thin cooling function for solar abundances.
Solid line is the result of our code, while dashed line is the result found by Dopita and Sutherland \cite{ds9}.
}
\label{Fig1}
\end{figure}

\subsection{Heating rates}

The dissipative dark matter halos not only cool but are also heated, with the heat source
originating from 
ordinary Type II supernovae. The assumed mechanism requires 
a significant fraction of a supernova's core collapse energy
to be converted, ultimately, into dark radiation. 
The kinetic mixing interaction plays an important role as it is responsible for the transfer
of energy to the mirror sector.
At the particle physics level,
kinetic mixing imbues the mirror electron and mirror positron with a 
tiny ordinary electric charge of magnitude $\epsilon e$.  
This enables particle processes leading to the production of light mirror 
particles: $e', \ \bar e', \ \gamma'$, 
to readily occur in the hot dense core of ordinary supernovae 
(e.g. 
plasmon decay to $e' \bar e'$, 
$e\bar e \to e'\bar e'$, $e' \bar e' \to \gamma' \gamma'$ etc.). 
The light mirror particles interact weakly enough with ordinary matter so that 
they can escape from the supernova core
and also from the collapsing star.\footnote{The interactions of the escaping 
mirror particles with the baryonic matter, though quite weak,
could still transfer a substantial amount of energy to the baryons.
It has been speculated \cite{lon} that this mechanism might 
facilitate the explosion of a  supernova,
as the transfer of energy via the escaping 
neutrinos may be inadequate, although there
is still much debate in the literature \cite{yyy}.
}

The rate at which the core collapse energy is transferred to light mirror particles can be 
estimated from \cite{raffelt,raffelt2}. This energy loss rate is given by:
\begin{eqnarray}
Q_P = \frac{8 \zeta_3}{9 \pi^3}  \epsilon^2 \alpha^2 \left(
\mu^2_e + \frac{\pi^2 T_{SN}^2
}{3}\right) T_{SN}^3 Q_1
\label{raf1x}
\end{eqnarray}
where $Q_1$ is a factor of order unity, $\mu_e$ is the electron chemical potential, and 
$T_{SN} \sim 30$ 
MeV is the temperature of the supernova core.
The observation of around a dozen neutrino interactions associated with SN1987A
\cite{sn1,sn2} suggests that
$Q_P$ should not exceed the energy loss rate due to neutrino emission.
This indicates a rough upper limit on $\epsilon$ of
around $\epsilon \lesssim  10^{-9}$ \cite{raffelt}.

Supernovae can provide a rather substantial energy source
if $\epsilon$ does indeed have a value near this upper limit. 
This energy is initially distributed among the various light mirror particles:
$e', \bar e', \gamma'$ (potentially also some fraction in $\nu'$).
These particles, injected into the region around a supernova, would undergo a variety
of complex processes, shocks etc. 
In this paper we follow previous work
\cite{foot13,foot14} and assume that the bulk of 
this energy is (ultimately) converted into $\gamma'$ emission.
These mirror photons, with total energy up to around half the supernova core collapse
energy ($\sim 10^{53} \ {\rm erg}$ per supernova), can propagate out into the halo. 
These photons can be absorbed via the dark photoionization process.
The key idea is that such mirror-photon heating, powered by ordinary supernovae, 
can replace  the energy dissipated in
the halo due to the various cooling processes.


If ordinary supernovae are the source of the heating 
of the halo, then we will need to know  their rate ($R_{SN}$)
and spatial distribution in a given galaxy.
Supernovae are the final evolutionary stage of large stars. 
Ultraviolet radiation is directly emitted from the photospheres of large stars 
with $M_* \gtrsim  3m_\odot$ (O- through later-type B-stars). 
Galactic UV flux measurements, such as those taken 
by the Galaxy Evolution Explorer (GALEX) satellite \cite{galex}, 
can therefore be used 
to probe the recent star formation rate over a timescale $\sim 100$ Myr
e.g. \cite{uv1,uv2,uv3}. 
It follows that a galaxy's UV luminosity, $L_{FUV}$, 
(taken here to be the far UV bandpass of 1350-1750 \AA), 
should provide an estimate of the 
current supernova rate.
That is, we expect the rough scaling:
$R_{SN} \propto L_{FUV}$. Using $L_{FUV} \propto 10^{-0.4 M_{FUV}}$, where $M_{FUV}$ is the galaxy's
FUV absolute magnitude,
we therefore expect:
\begin{eqnarray}
R_{SN} \approx R_{SN}^{MW}  \ \frac{10^{-0.4M_{FUV}}}{10^{-0.4M_{FUV}^{MW}}}
\ .
\end{eqnarray}
Here, $M_{FUV}^{MW}$
is the FUV absolute magnitude  
for the Milky Way, and 
$R_{SN}^{MW} \sim 10^{-9} \ {\rm s}^{-1}$ 
is the Type II supernova rate in the Milky Way.
[We take $M_{FUV}^{MW} =  -18.4$
in the numerical work.]

Supernovae become a powerful source of mirror photons with uncertain spectrum and total energy.
We denote the
average dark photon luminosity of a single supernova by
$L_{SN}$.
The frequency spectrum will be modelled, for simplicity, by a 
thermal distribution:
\begin{eqnarray}
\frac{dL_{SN}}{dE_\gamma} = \frac{15}{\pi^4 T^4_{eff}} \ \frac{E_\gamma^3}{e^{E_\gamma/T_{eff}} - 1}
\ L_{SN} 
\ .
\label{isis}
\end{eqnarray}
The relevant effective temperature parameter, $T_{eff}$, is not known. 
One would need to be able to model the complex processes in the expanding $e',\bar e',\gamma'$
plasma around a supernova. In the absence of such modelling, we
consider a wide range of potential $T_{eff}$ values.

A thermal distribution for the supernova sourced spectrum of dark photons
is very convenient, but may be a poor representation of
the actual spectrum. In fact, this system may have some similarities with gamma ray bursts,
which are very complex, and are seldom thermal.
Modelling the system with an
alternative distribution,
e.g. a power law, would appear to be equally valid given this state of ignorance.

Consider now the supernova spatial distribution within a given galaxy. For rotationally 
supported galaxies, this  distribution 
could be modelled as a Freeman disk (located at $\theta = \pi/2$ in the spherical coordinate 
system with origin at the
galactic center), so that the differential source luminosity of SN dark photons takes the form:
\begin{eqnarray}
\frac{d{\cal L}^S_{SN}({\bf r})}{dVdE_\gamma} =
\frac{e^{-r/r_D}}{2\pi r_D^2 r} 
\frac{dL_{SN}}{dE_\gamma} 
R_{SN}
\delta(\theta - \pi/2) 
\ .
\end{eqnarray}
Since we are solving for the steady state configuration assuming spherical symmetry,
we shall replace this supernova disk distribution with a spherically symmetric
analogue of the form [cf. discussion around Eq.(\ref{doc})]:
\begin{eqnarray}
\frac{d{\cal L}^S_{SN}(r)}{dVdE_\gamma} =
\frac{e^{-r/r_D}}{4\pi r_D^2 r}  
\frac{dL_{SN}}{dE_\gamma} 
R_{SN}
\ .
\label{37yy}
\end{eqnarray}
For the Milky Way, we have a rough upper limit:  $R_{SN}^{MW}L_{SN} \lesssim 10^{45}\ $ erg/s.

The differential source luminosity,
at a given location,
is the sum of the radiation cooling function [Eq.(\ref{wed1})] 
and the SN sourced photons, i.e.
\begin{eqnarray}
\frac{d{\cal L}^S(r)}{dVdE_\gamma} 
= \frac{d{\cal C}(r)}{dE_\gamma} + 
\frac{d{\cal L}^S_{SN}(r)}{dVdE_\gamma}
\ .
\end{eqnarray}
Recall that the above source luminosity is required to compute the
differential flux at a given location [$dF(r)/dE_\gamma$]  
via Eq.(\ref{flux1}), and that this
flux is needed to compute the ionization state [Eq.(\ref{trump})]. 
The flux is also required to calculate
the differential rate of radiation absorption (heating rate):
\begin{eqnarray}
\frac{dH(r)}{dE_\gamma}
 = \sum_{A,k} \ \sigma[A^k]_{PI} \ n_{A^k} \  E_\gamma \ \frac{dF(r)}{dE_\gamma}  
\ .
\label{sfu1}
\end{eqnarray}
We now have a set of interconnected equations describing the ionization state, cooling and heating
rates of a dark plasma. These equations will need to be solved 
to find the steady state solution for mirror dark matter galaxy halos.
These equations, with straightforward
modifications (as indicated), are applicable also to the more generic dissipative model of section 2.1.


\section{Steady state solution}

\subsection{The numerical method}

The system of equations governing the ionization state, the heating and cooling rates, are somewhat 
nontrivial to solve. Our strategy to solve them
is to choose a suitable form for the density profile (defined in terms of several parameters
to be determined from the dynamics). The system of equations is then solved iteratively as follows: 
\vskip 0.3cm
\noindent
(a) With the chosen density profile, 
the temperature profile is calculated from the hydrostatic
equilibrium condition [Eq.(\ref{eqmx})] (in the first iteration the 
$\bar m$ profile is chosen arbitrarily, in
the second and subsequent iterations it is 
input from the previous iteration).
\vskip 0.3cm
\noindent
(b) Using the temperature profile calculated from step (a) the ionization state 
is computed
[Eq.(\ref{trump})],
and a new $\bar m$ profile derived
(in the first iteration the mirror photon flux can be neglected,
in the second and subsequent iterations
the flux is input from the previous iteration).
\vskip 0.3cm
\noindent
(c) Using the results from steps (a) and (b),  the  
heating [${\cal H}(r)$], cooling [${\cal C}(r)$] rates 
are evaluated [from Eq.(\ref{sfu1}) and Eq.(\ref{wed1})], 
and also the mirror photon flux $dF(r)/dE_\gamma$ [from Eq.(\ref{flux1})].
\vskip 0.3cm
\noindent
The above three steps can be repeated until a 
stable solution for $\cal H$ and $\cal C$ emerges (typically
requires around 10-20 iterations).
If the chosen density profile is such that ${\cal H} \simeq {\cal C}$ 
at each location, then this density profile (together with the temperature 
derived from the hydrostatic equilibrium
condition via the above iterative procedure) 
would represent an approximate steady state solution to the fluid equations.

We have made use of two spherically symmetric dark matter halo 
density profiles. The first one is the density profile that arises in the idealized case
of an isothermal halo in the  optically thin limit  \cite{foot16,olga}.
Under these assumptions,
${\cal C}({\bf r}) \propto n({\bf r})^2$ and ${\cal H}({\bf r}) \propto n({\bf r}) F({\bf r})$ 
(where $F({\bf r})$ is the 
flux of dark photons at ${\bf r}$), the steady state condition ${\cal C}({\bf r}) = {\cal H}({\bf r})$ 
implies that $n({\bf r}) \propto F({\bf r})$. For a flux originating from a spherical distribution 
of SN heat sources, this yields a dark matter density profile of the form:
\begin{eqnarray}
\rho(r) = \lambda \int \int \frac{d{\cal L}^S_{SN}(r')}{dV} \ 
{\cal F}(r,r',\theta') \ 2\pi r'^2 d\cos\theta' dr'
\label{r1}
\end{eqnarray}
where ${\cal F}(r,r',\theta') = 1/(4\pi [r^2 + r'^2 - 2rr'\cos\theta'])$. Also, from Eq.(\ref{37yy}) we have:
\begin{eqnarray}
\frac{d{\cal L}^S_{SN}(r)}{dV} =
\frac{\kappa e^{-r/r_D}}{4\pi r_D^2 r}  
\end{eqnarray}
where $\kappa \equiv R_{SN}L_{SN}$.
The coefficient $\lambda$ would be independent of $r$ in
the optically thin and isothermal limit.
We refer to this one-parameter distribution as the $\lambda$-density profile.
We have also considered a generic cored profile,
\begin{eqnarray}
\rho  (r) = \rho_0 \left[ \frac{r_0^2}{r^2 + r_0^2}\right]^{\beta}
\ .
\label{r2}
\end{eqnarray}
This profile is defined in terms of three independent parameters: $\rho_0, \ r_0,\ \beta$.
With either of these profiles we can follow the steps (a)-(c) iterated until a 
stable solution for $\cal H$ and $\cal C$ emerges.

For $\rho(r)$ to be an approximate steady state solution requires  
${\cal H} \simeq {\cal C}$ at every location in the halo.
To quantify this, it is  useful to 
introduce the functional $\Delta$:
\begin{eqnarray}
\Delta \equiv \frac{1}{R_2-R_1} \int^{R_2}_{R_1} 
\frac{|{\cal H}(r') - {\cal C}(r')|}{{\cal H}(r')+{\cal C}(r')} \ dr' 
\end{eqnarray}
where we take $R_1 = 0.3r_D$, $R_2 = 10r_D$ in our numerical work.
We then minimize $\Delta$ with respect 
to variations in $\lambda$ for the $\lambda$-density profile, and $\rho_0, \ r_0, \ \beta$,
for the generic cored profile.
If this minimum is sufficiently small, say less than 0.05, then 
we shall suppose that a candidate steady state solution exists.
The value of $\lambda$ (or  $\rho_0, \ r_0,\ \beta$) 
that minimizes $\Delta$ defines the density profile
of the candidate solution.

\subsection{Mirror dark matter}

Following the iterative procedure outlined above, we have
searched for steady state solutions for mirror dark matter halos with realistic asymptotic 
halo velocity for a Milky Way scale galaxy ($v_{rot}^{asym} \approx 200$ km/s). 
We explored a wide range of
the available parameters including the halo metal abundance parameter, $-2.0 \le \zeta
\le 2.0$,
and SN parameters: $T_{eff} \le 1000$ keV, 
$\kappa \le 10^{46} \ {\rm erg/s}$.
We also looked at modifications of 
the halo metal composition, e.g. $Fe'/O'$ ratio etc.,
and different forms for the SN spectrum, e.g. replacing the thermal spectrum, Eq.(\ref{isis}), with
a power law. Throughout this parameter space it was found that halo cooling exceeds heating.  
[However, for a limited parameter space, $\zeta \sim 2.0, \ 
T_{eff} \sim 1 \ {\rm keV}, \ \kappa \sim 10^{46} \
{\rm erg/s}$, we found that cooling only exceeds heating by factor of $\sim 3$.] 
That is, we were unable to find a steady state solution for mirror dark matter  galactic halos 
with realistic halo density.  
Previous more optimistic results of \cite{foot13,foot14} were due, in part, to the incomplete treatment
of cooling (neglect of line emission and recombination radiation), and, in part, to the incomplete
treatment  of the ionization state (neglect of the photoionization
contribution).


\begin{figure}[t]
  \begin{minipage}[b]{0.5\linewidth}
    \centering
    \includegraphics[width=0.7\linewidth,angle=270]{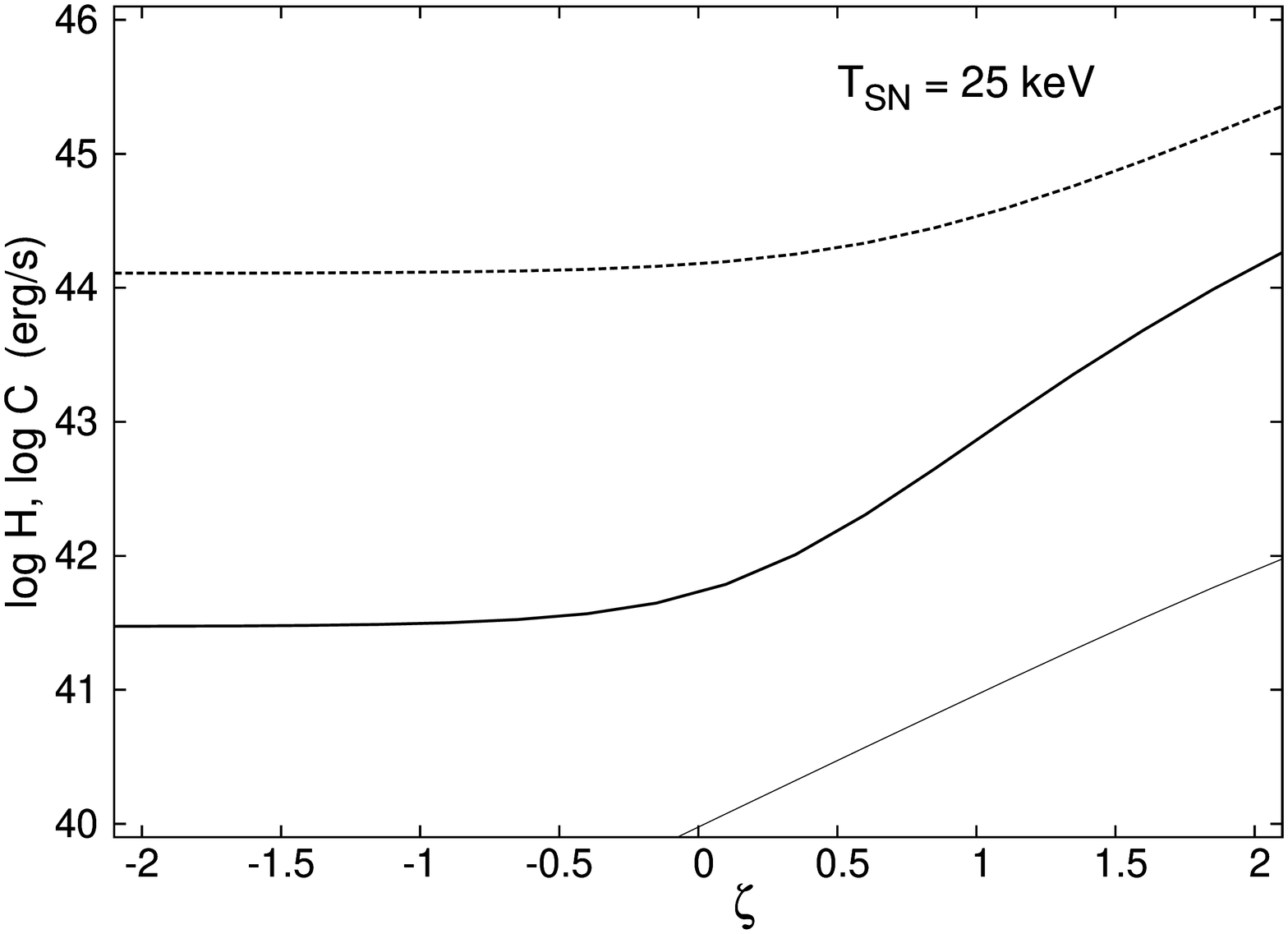}
     (a)
    \vspace{4ex}
  \end{minipage}
  \begin{minipage}[b]{0.5\linewidth}
    \centering
    \includegraphics[width=0.7\linewidth,angle=270]{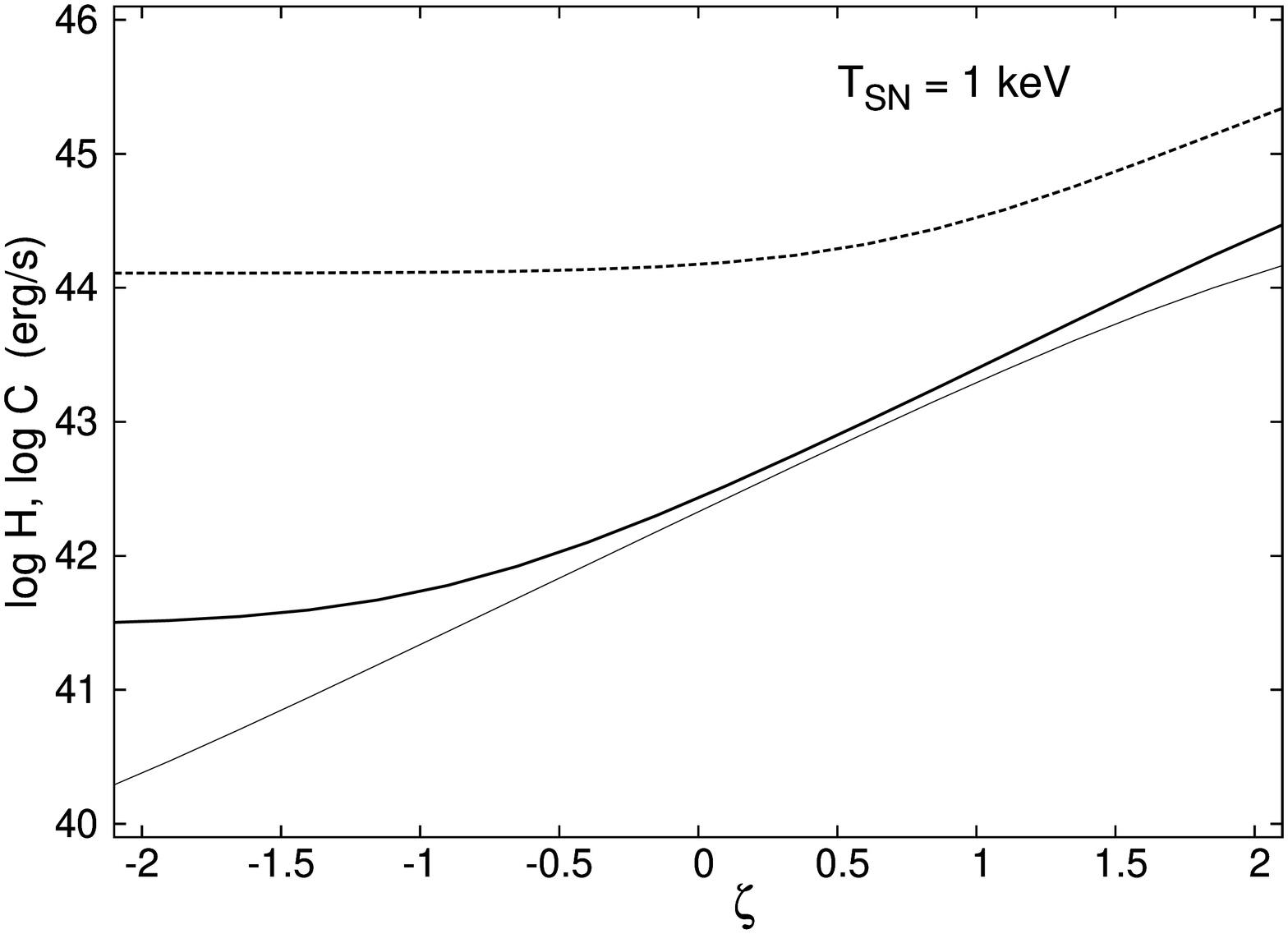}
    (b)
    \vspace{4ex}
  \end{minipage}
  \begin{minipage}[b]{0.5\linewidth}
    \centering
    \includegraphics[width=0.7\linewidth,angle=270]{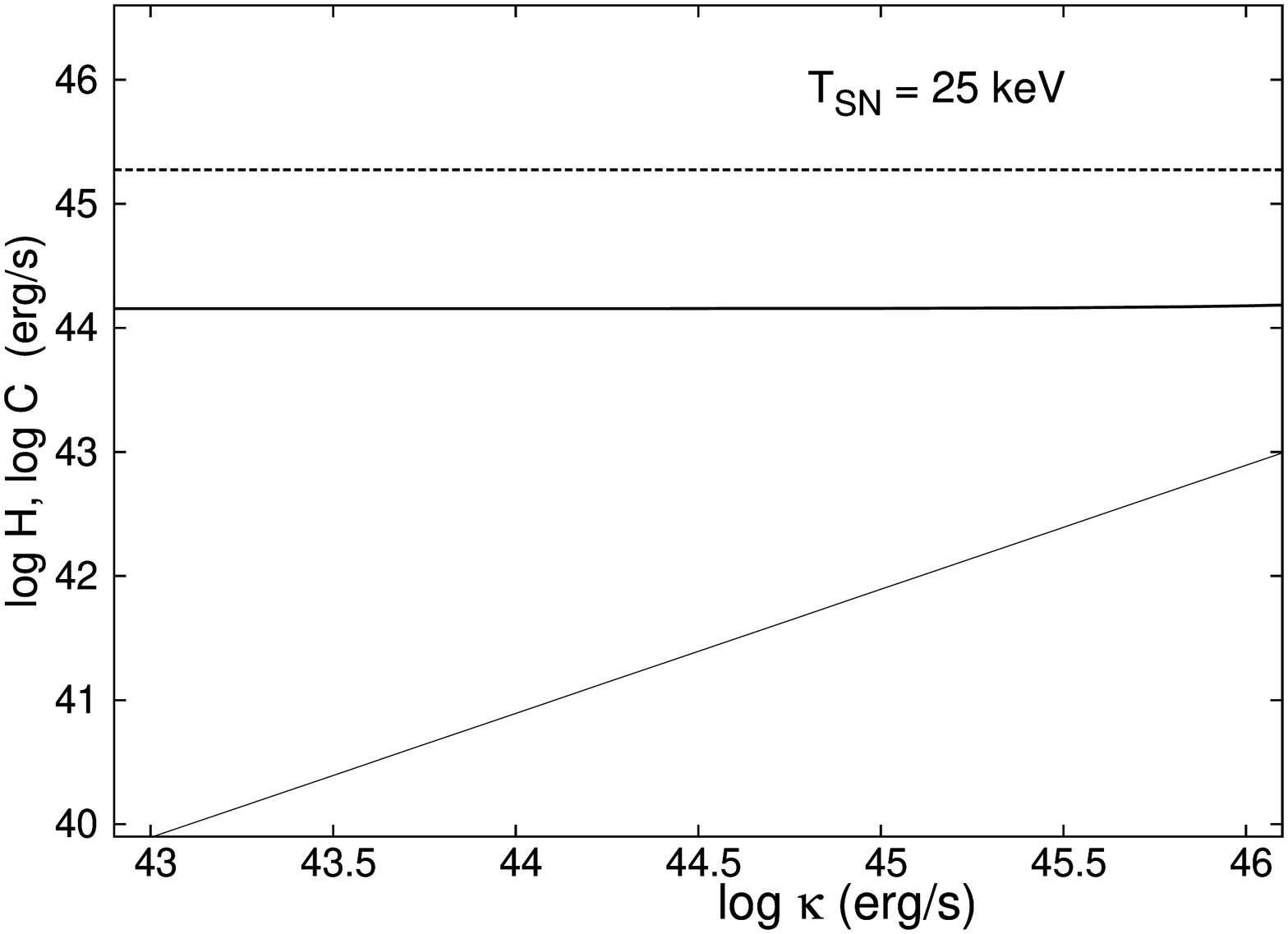}
     (c)
    \vspace{4ex}
  \end{minipage}
  \begin{minipage}[b]{0.5\linewidth}
    \centering
    \includegraphics[width=0.7\linewidth,angle=270]{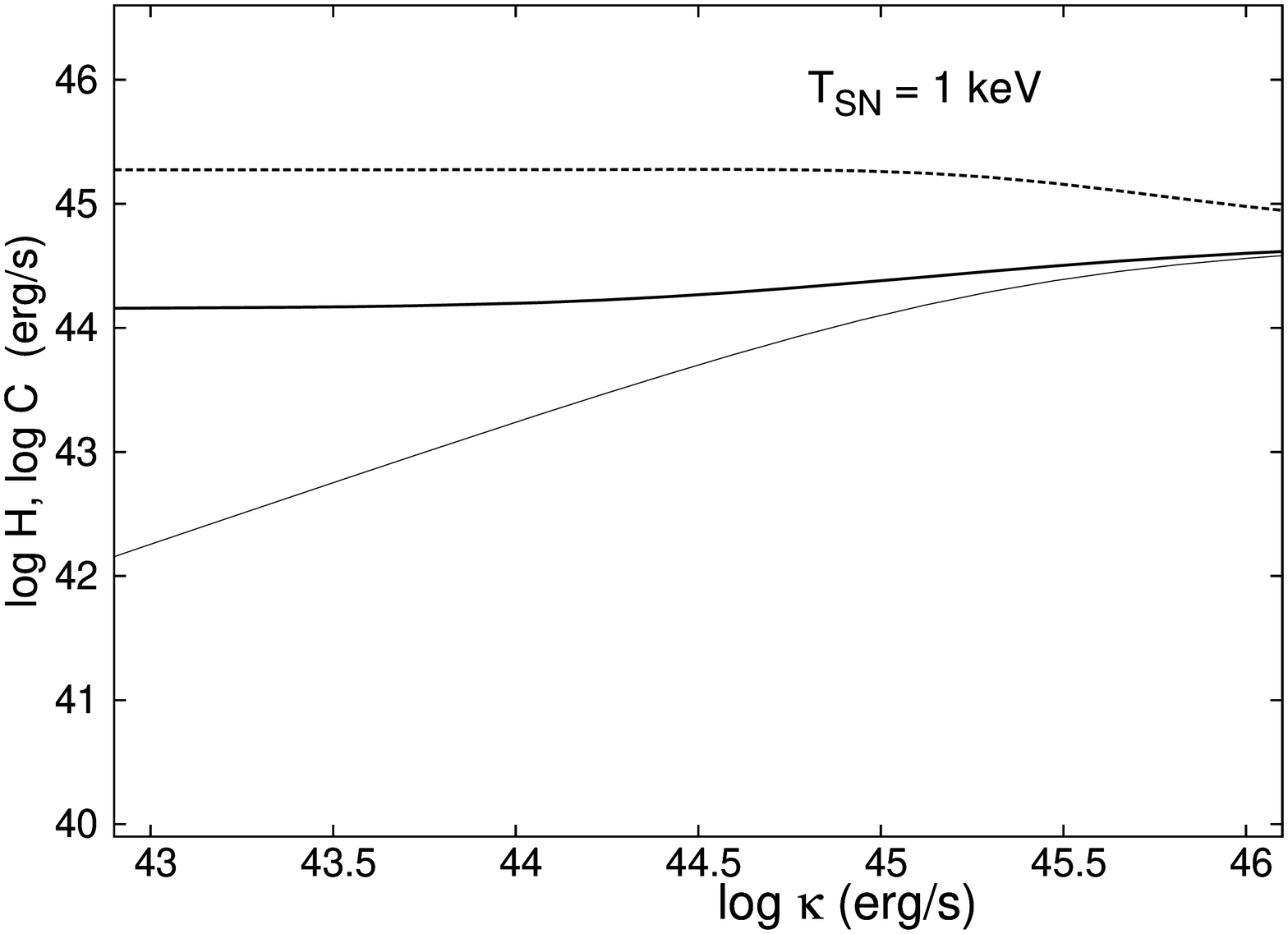}
    (d)
    \vspace{4ex}
  \end{minipage}
\vskip -1.0cm
\caption{\small 
Halo heating (thick solid line) and cooling rates (dashed line) 
for a Milky Way scale galaxy. The thin solid line is the heating contributed by SN sourced dark photons.
The $\lambda-$density profile was used
with the halo temperature and ionization state numerically determined from
steady state equations.
(a)  and (b) give the rates in terms of the mirror metal abundance parameter, $\zeta$,
for $\kappa = 10^{45}$ erg/s and (a) $T_{SN} = 25$ keV, (b) $T_{SN} = 1$ keV. 
While (c) and (d) give the rates in terms of the SN sourced
dark photon luminosity, $\kappa$ 
[erg/s], 
for $\zeta = 2$ and (c) $T_{SN} = 25$ keV, (d) $T_{SN} = 1$ keV.
}
\end{figure}

In Figure 2, we illustrate the problem by showing the integrated heating and cooling
rates ($H = \int {\cal H} dV, \ C = \int {\cal C} dV$) 
for a Milky Way scale galaxy. We considered the $\lambda-$density profile with
$\lambda \kappa = 7.3\times 10^{9}  \ [m_\odot/{\rm kpc}]$, a parameter choice
sufficient to give a  realistic asymptotic halo rotation velocity of $\sim 220$ km/s. With this 
density profile, the halo temperature  and ionization state  were 
determined  at each location in the halo from the steady state equations 
using the iterative numerical method outlined earlier.
Figure 2 clearly indicates that the halo
cooling always exceeds heating for the range of 
$\zeta$, $T_{SN}$, $\kappa$ parameters examined. 
With the large values of $\kappa \equiv R_{SN}L_{SN}$ considered, there is enough energy available, the problem is that 
this energy is not readily absorbed in the halo. To illustrate this issue, we show in Figure 3 the 
optical depth for a dark photon
originating near the galactic center and  escaping the galaxy (again for a Milky Way scale galaxy with halo properties as per Figure 2).
As this figure shows, the optical depth is typically less than unity.
Increasing the supernovae rate beyond observational limits does not help as the increased
energy in dark photons is compensated by a reduction in the optical depth caused by the increased
ionization.


\begin{figure}[t]
  \begin{minipage}[b]{0.5\linewidth}
    \centering
    \includegraphics[width=0.7\linewidth,angle=270]{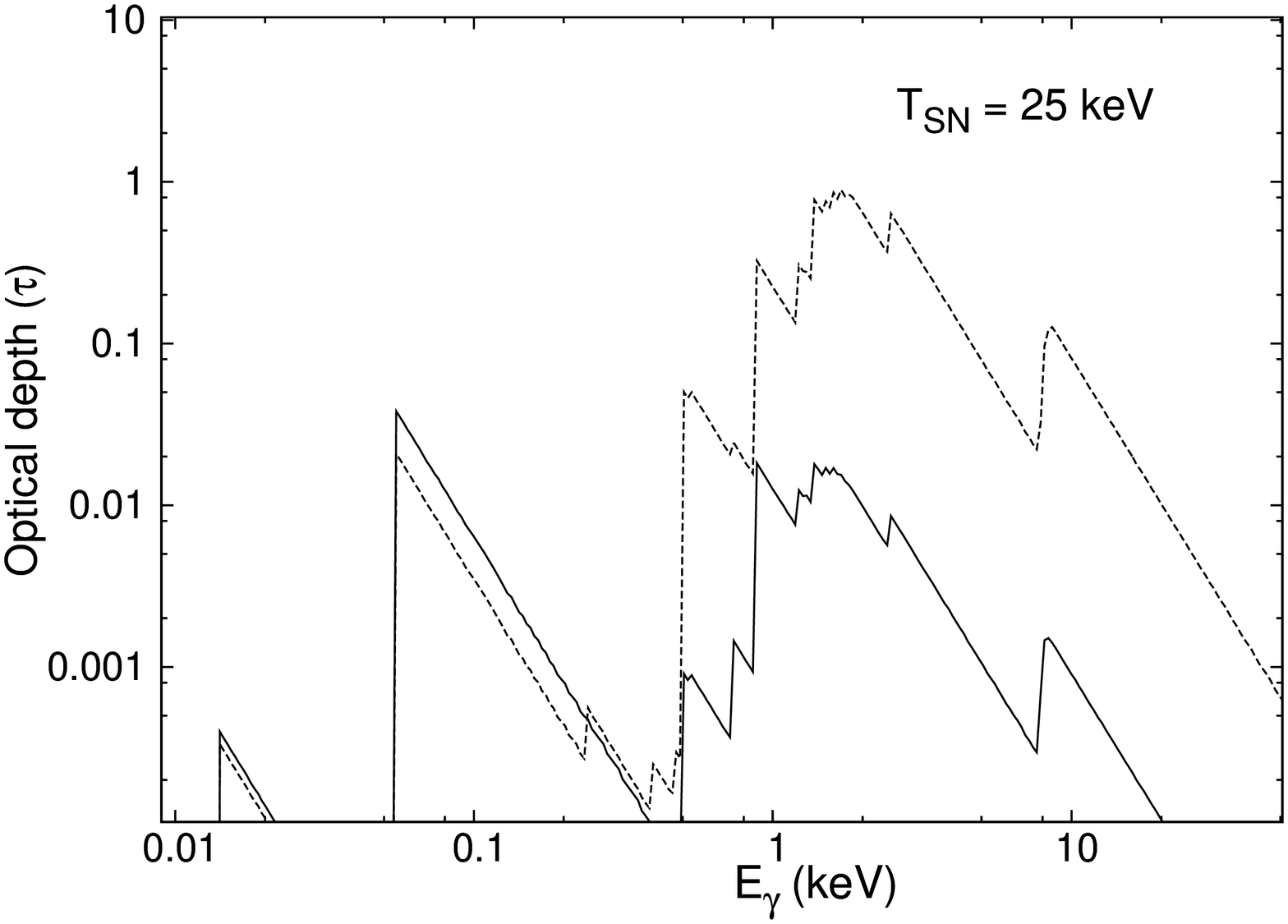}
     (a)
    \vspace{4ex}
  \end{minipage}
  \begin{minipage}[b]{0.5\linewidth}
    \centering
    \includegraphics[width=0.7\linewidth,angle=270]{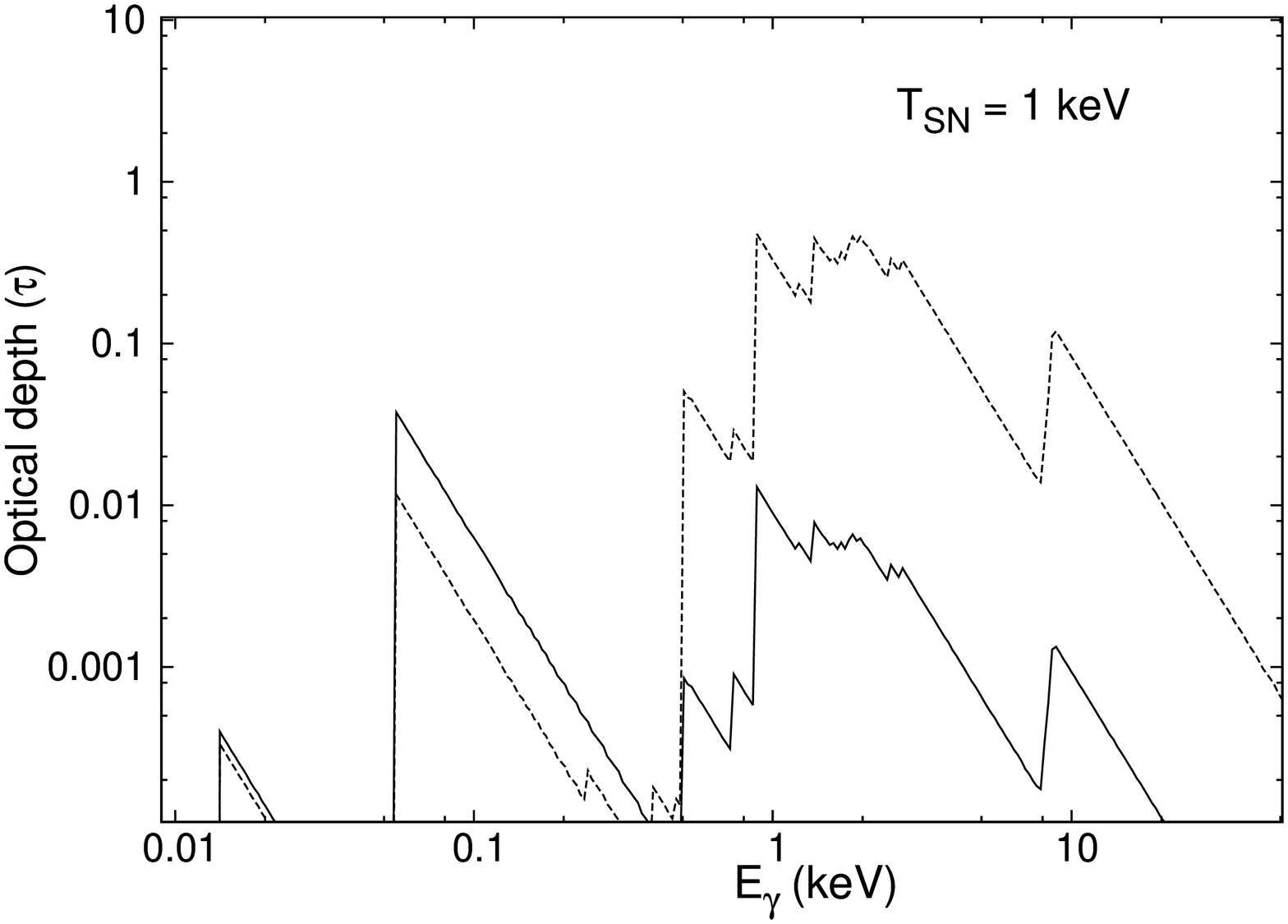}
    (b)
    \vspace{4ex}
  \end{minipage}
\vskip -1.0cm
\caption{\small 
The optical depth for a dark photon
originating near the galactic center and  escaping a Milky Way scale galaxy, 
with halo properties as per Figure 2.
The solid (dashed) line shows results for $\zeta = 0$ ($\zeta = 2$),
with $\kappa = 10^{45}$ erg/s and (a) $T_{SN} = 25$ keV,  
(b) $T_{SN} = 1$ keV.  
}
\end{figure}

The discouraging results reported here for mirror dark matter 
might be due to residual simplifications, or possibly, invalid assumptions. 
The simplifications include:
treating the mirror ions and electrons
as a single-component fluid with a common local temperature $T({\bf r})$,
simplified description of SN sourced dark photon energy spectrum,
neglect of dark magnetic fields etc.
One of the questionable assumptions involves the mechanism by which the
SN sourced energy is transported to the halo.
In this paper, we have followed earlier work \cite{foot13,foot14} and 
assumed that this energy is transmitted to the halo
via dark radiation.  It is possible that the SN sourced energy 
is instead transported to the halo via local collisional processes in the SN vicinity.
A discussion of this alternative energy source will be given in a separate 
paper.
In any case, the above caveats
suggest that  no definite conclusion as to the validity, or otherwise, 
of the mirror dark matter model
could be made at this time; 
in the remaining discussion of this paper we shall focus on
alternative dissipative particle models.

\subsection{Two-component dissipative model}

The mirror model is rather unique in that the fundamental interaction cross sections have
no free parameters: they are all identical to those of the 
corresponding ordinary particle processes.
Naturally, it is also worthwhile to look at more generic dissipative models, the simplest
such model is the two-component model of \cite{sunny1}, reviewed in section 2.1. In that model
the dark halo consists of just two matter components, the dark electron and dark proton
(with dark charge ratio: $Z' \equiv |Q'(p_d)/Q'(e_d)|$).
That model has five fundamental parameters: $m_{e_d}, m_{p_d}, Z', \alpha_d$ and
$\epsilon$. The considered parameter space is somewhat restricted: $m_{e_d} \ll m_{p_d}$ and $Z' \ge 1$, so that atoms could 
potentially form with a $p_d$ nucleus surrounded by 1
or more dark electrons.

For such a dissipative dark matter model to have the potential of being realistic,
galactic halos should not be fully ionized. 
This is required so that the halo can absorb the supernova sourced dark photons
via the photoionization process.
This restriction leads to the rough criterion: $T_{halo} \lesssim I_{1}$, 
where $I_1$ is the binding energy of 
the inner most (K shell) dark electron. 
This condition is most restrictive for the largest galaxies, and can be used to estimate  
an upper bound on $m_{p_d}$.
The halo also needs to have a non-negligible  degree of ionization, even for the smallest galaxies.
Under the assumption that the
ionization is due primarily to dark electron scattering, this criterion leads to a  
lower bound on $m_{p_d}$. (It might be possible to weaken this lower bound given the photoionization
contribution, and further study could be done to clarify this issue.)
These conditions, derived in Eq.(91) of \cite{sunny1}, imply that
$Z' \ge 3$ and that $m_{p_d}$ lies in the range:
\begin{eqnarray}
\left(\frac{Z'}{10} \right)\left(\frac{\alpha_d}{10^{-2}} \right)^2 \left( \frac{m_{e_d}}{\rm MeV} \right) 
\lesssim \frac{m_{p_d}}{{\rm GeV}} \lesssim 
100 \left( \frac{Z'}{10} \right)^3\left(\frac{\alpha_d}{10 ^{-2}} \right)^2 \left( \frac{m _{e_d}}{\rm MeV}
\right) g(\alpha_d,Z')
\label{mf2bounds}
\end{eqnarray}
where $g(\alpha_d,Z') \equiv {\rm max} (\alpha_d^3 Z'^4, 1)$.

For a given choice of SN parameters (we take $T_{eff} = 25$ keV, $\kappa   = 10^{45}$  
erg/s for definiteness), we have searched 
for $m_{p_d}, \ m_{e_d}$ and $\alpha_d$ values which give realistic asymptotic rotational
velocity for a Milky Way scale galaxy (we fixed $Z' = 6$ for definiteness).
There is a significant parameter space where this occurs, and we shall focus here 
on a specific example:
\begin{eqnarray}
m_{p_d} = 100m_p, \ m_{e_d} = 8m_e, \ \alpha_d = 4\alpha, \ Z'=6
\ .
\end{eqnarray}
For the particular dissipative dark matter model
defined by these parameters,
we have undertaken a search for steady state solutions 
for a representative range of galaxies.

\begin{figure}[t]
  \begin{minipage}[b]{0.5\linewidth}
    \centering
    \includegraphics[width=0.7\linewidth,angle=270]{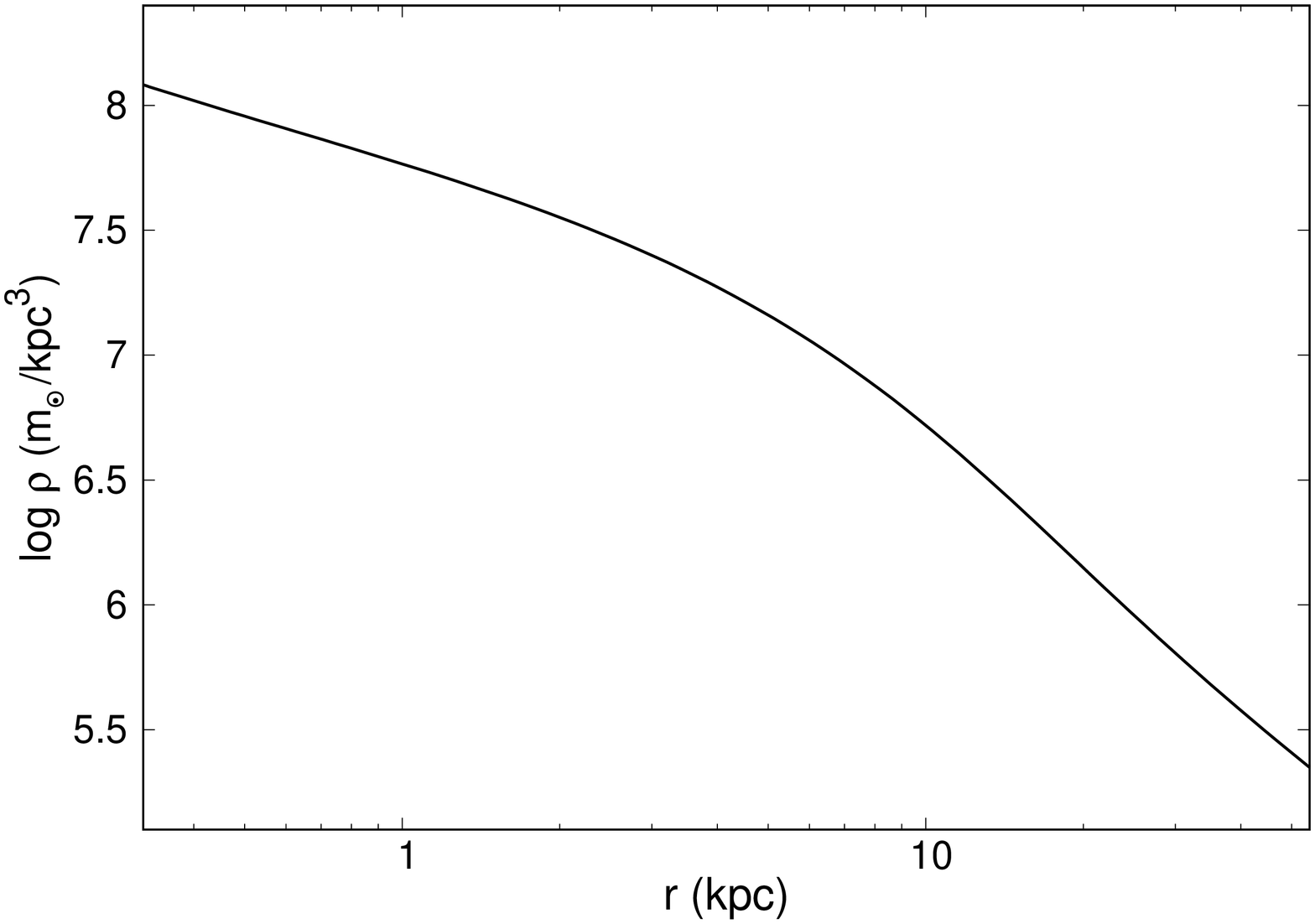}
     (a)
    \vspace{4ex}
  \end{minipage}
  \begin{minipage}[b]{0.5\linewidth}
    \centering
    \includegraphics[width=0.7\linewidth,angle=270]{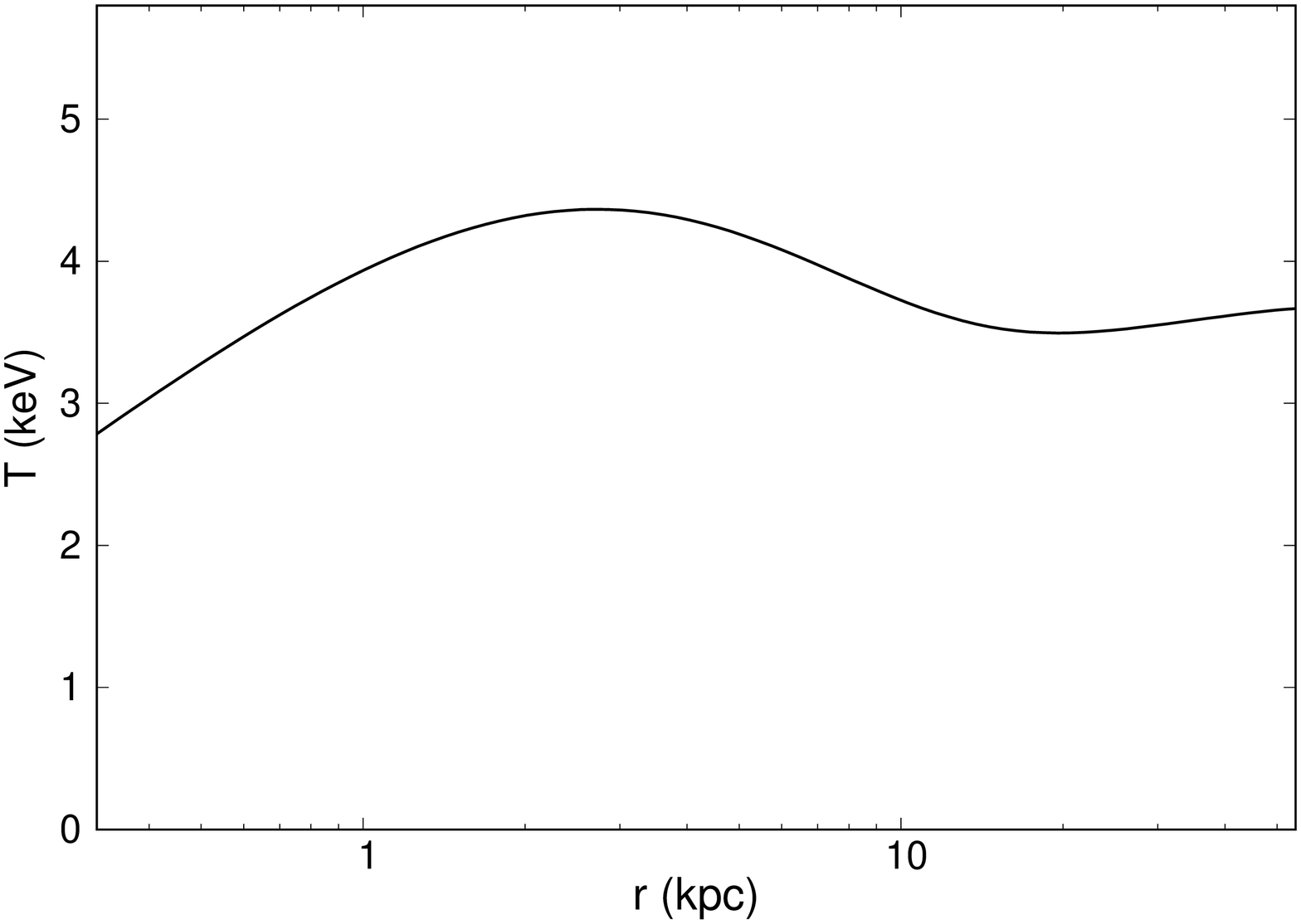}
    (b)
    \vspace{4ex}
  \end{minipage}
  \begin{minipage}[b]{0.5\linewidth}
    \centering
    \includegraphics[width=0.7\linewidth,angle=270]{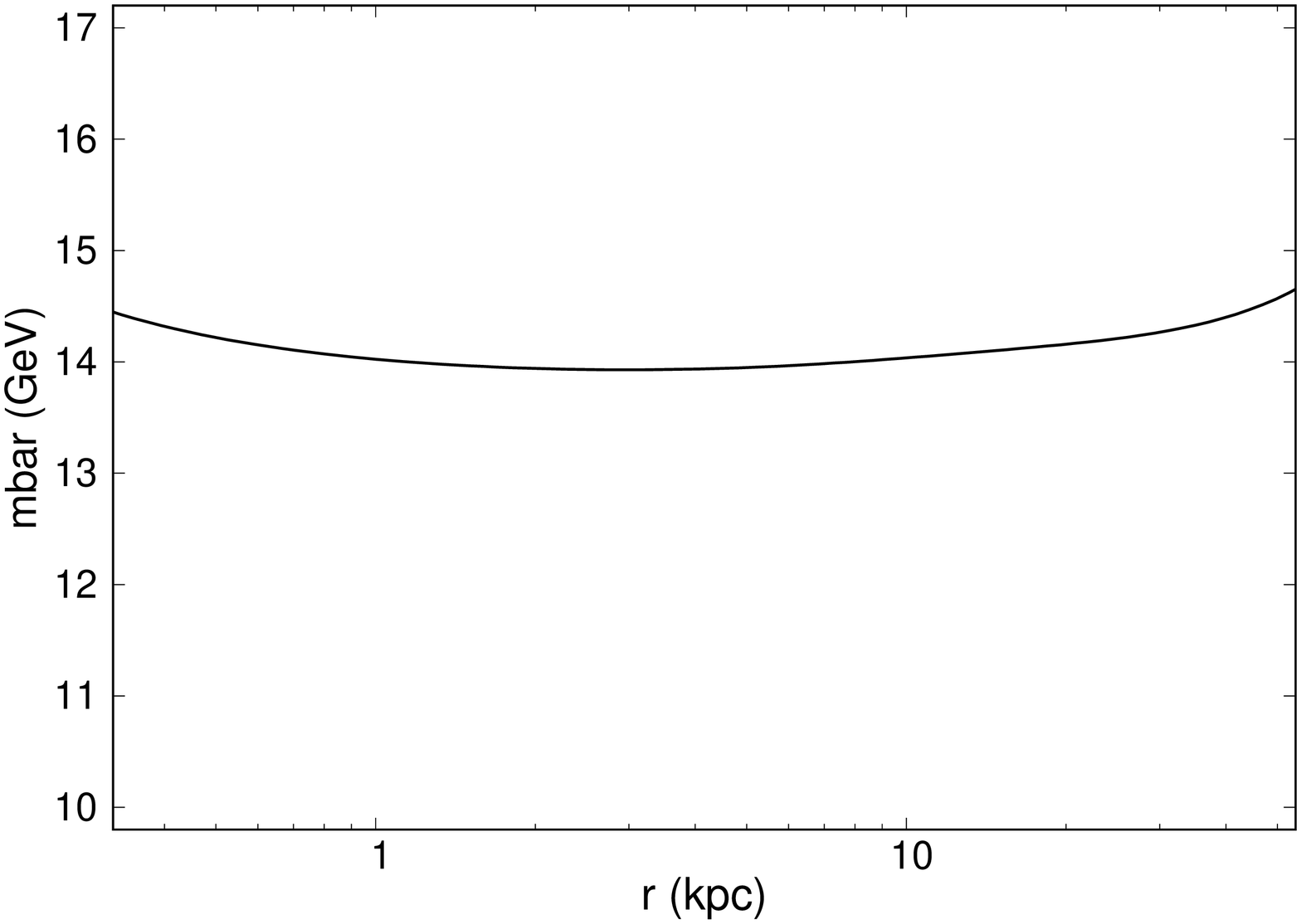}
     (c)
    \vspace{4ex}
  \end{minipage}
  \begin{minipage}[b]{0.5\linewidth}
    \centering
    \includegraphics[width=0.7\linewidth,angle=270]{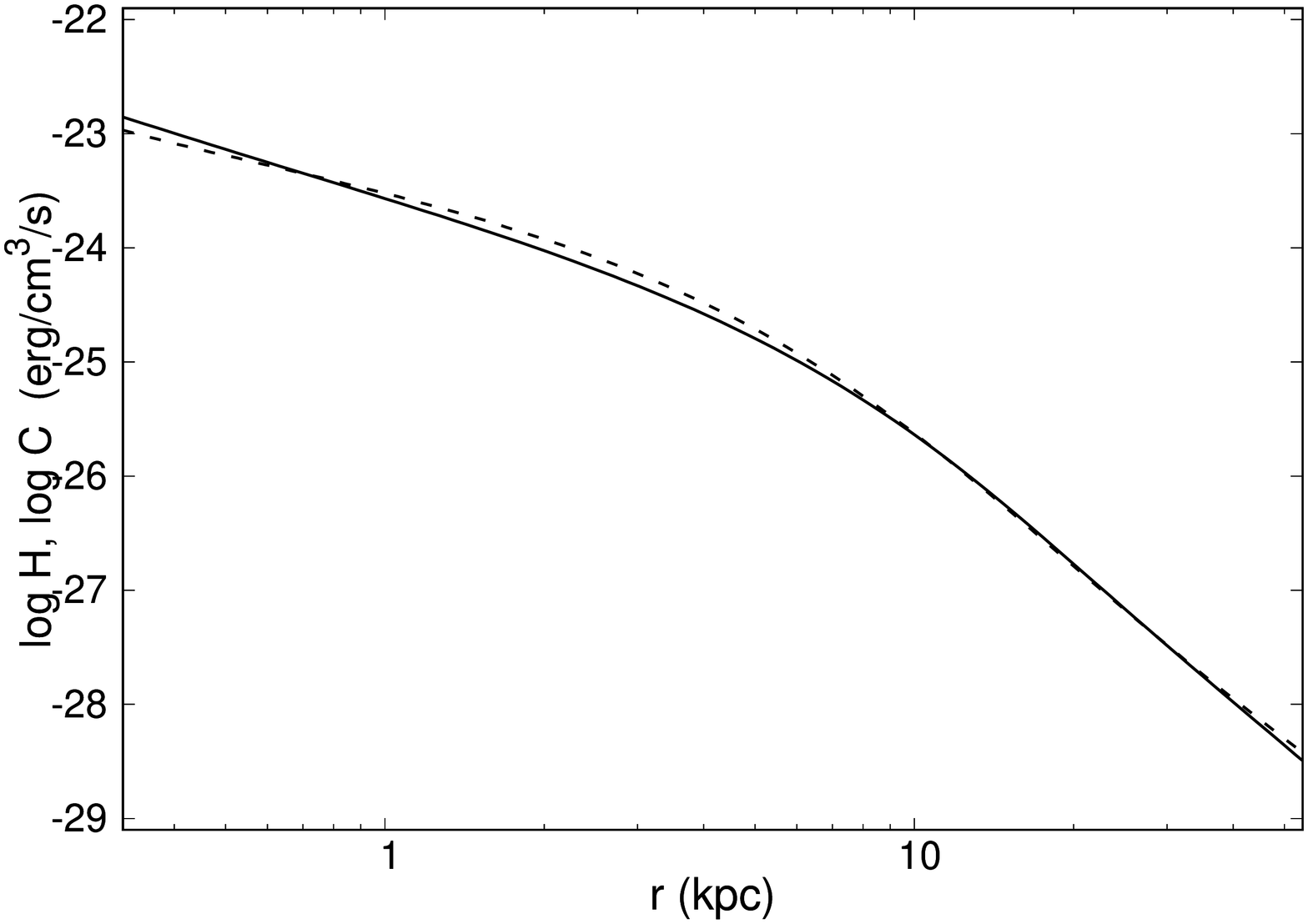}
    (d)
    \vspace{4ex}
  \end{minipage}
\vskip -1.0cm
\caption{\small Properties of the  steady state solution  obtained for a
Milk Way scale galaxy ($m_{baryons} = 10^{11} \ m_\odot$, $f_s = 0.8$, 
$r_D = 3.95$ kpc, $M_{FUV} = -18.4$).
Shown are (a) the
halo density, (b) the halo temperature, (c) the mean mass, $\bar m$,  
and (d) the heating, cooling rates [${\cal H}, \ {\cal C}$]  
(solid, dashed line).  
}
\end{figure}

Consider first
a Milky Way scale galaxy with baryonic parameters:
$m_{baryons} = 10^{11} \ m_\odot$, $f_s = 0.8$, $r_D = 3.95$ kpc, $M_{FUV} = -18.4$.
The system of equations describing the ionization state, as well as the 
heating and cooling rates, were solved iteratively
with the $\lambda$-density profile [Eq.(\ref{r1})]. An approximate
steady state solution was identified 
for $\lambda = 7.3\times 10^{-36}  [m_\odot/{\rm kpc} \ {\rm s/erg}]$, with 
$\Delta_{\rm min} \simeq 0.05$. 
To better represent this putative solution, we replaced the
coefficient, $\lambda$, in Eq.(\ref{r1}),  with the radial
expansion:
\begin{eqnarray}
\lambda \to \lambda \left[1 \ + \ \sum_{n=1}^{N} a_n \left(\frac{r}{r_D}\right)^n 
 +  b_n  \left(\frac{r_D}{r}\right)^n\right]
\ .
\label{2grl}
\end{eqnarray}
In our numerical work we considered
only $N=1$ as this was  sufficient to significantly reduce the $\Delta_{\rm min}$ value
of the approximate steady state solution.
In Figure 4 we show some physical properties of the solution found.
Evidently,
the halo's temperature profile is nearly isothermal,
and  the density profile is close to quasi-isothermal (to be discussed in more detail
shortly).

\begin{table}[b]
\centering
\begin{tabular}{c c c c}
\hline\hline
$m_{baryons} (m_\odot)$ & $r_D$ (kpc) & $M_{FUV}$ & $f_s$  \\
\hline
$10^{11}$  & 3.95 & -18.4  &     0.8   \\
$10^{10.5}$  & 2.70  &  -17.9 &      0.8   \\
$10^{10}$  & 2.00 &   -17.4 &     0.8  \\
$10^{9.5}$  & 1.60  &  -16.9 &      0.8    \\
$5 10^{8}$  & 0.60  &  -15.0 &      0.2    \\
$10^{8}$  & 0.40  &  -13.4 &      0.2    \\
\hline\hline
\end{tabular}
\caption{\small Baryonic properties (baryon mass, baryonic scale length, FUV
absolute magnitude, and stellar mass fraction)
for the six `canonical' model galaxies considered.
}
\end{table}

The above procedure can be repeated for other model galaxies. One need only input
the baryonic parameters, $m_{baryons}, f_s, r_D, M_{FUV}$, and the halo properties
can be computed from the steady state condition.
We have examined large  stellar dominated galaxies  
(putative spiral galaxies)
with baryon masses: $10^{9.5}, \ 10^{10},\ 10^{10.5}, \ 10^{11} \ m_\odot$. The  stellar
mass component, with mass fraction  set to $f_s = 0.80$, was  
assumed to be distributed as in Eq.(\ref{doc}), with baryonic scale length values ($r_D$)
typical of high surface brightness spirals (taken from Eq.(8) of \cite{sal07}). 
The remaining baryon fraction ($1.0-f_s$), the gas component, 
was modelled with a more extended distribution ($r_D^{gas} = 3r_D$, as in e.g. \cite{sal17}). 
We also looked at
two small gas rich galaxies (putative dwarf irregular galaxies) 
with baryon masses $5\times 10^8 \ m_\odot$ and $10^8 \ m_\odot$ and 
with stellar mass fraction $f_s = 0.20$.
The FUV absolute magnitude values for this parameter set were chosen consistently with
the measured GALEX luminosities \cite{galex} of THINGS \cite{things} and 
LITTLE THINGS \cite{littlethings}
galaxies. 
For  these six examples we have numerically solved the system of equations iteratively
with the generalized $\lambda$-density profile of Eq.(\ref{r1}), Eq.(\ref{2grl}).
For all these examples approximate steady state solutions were 
found with $\Delta_{min} \approx 0.01-0.04$.
For ease of notation,
these approximate steady state solutions will be hereafter referred to as 
{\it steady state solutions}.
The galaxy baryonic parameters chosen  
are summarized in Table 2.


The one-parameter $\lambda$-density profile, Eq.(\ref{r1}),
represents a reasonable first order approximation to all of the steady 
state solutions found.
It is not surprising then, that much of what dissipative dark matter models predict can
be understood from the properties of that profile.
The $\lambda$-density profile has the asymptotic behaviour:
\begin{eqnarray}
\rho (r)  =
\frac{\lambda \kappa  }{4\pi r^2}   \ \ \ {\rm for} \ r \gg r_D\ ,
\label{bnb}
\end{eqnarray}
and rises logarithmically for $r \lesssim r_D$.
Over the finite range: $0.3 < r/r_D < 10$, 
the $\lambda$-density profile 
is numerically equivalent, to a good approximation (within $\sim$ few percent), 
to the density,
\begin{eqnarray}
\rho  (r) = \rho_0 \left[ \frac{r_0^2}{r^2 + r_0^2}\right]
\left[ 1 + ln \ \left( \frac{r^2 + r_0^2}{r^2}  \right) \right]
\label{r2z}
\end{eqnarray}
with
$r_0 = 1.75 r_D$ and $\rho_0 = 0.029 \lambda \kappa/r_D^2$.
This profile resembles the quasi-isothermal profile often adopted in the literature
to fit rotation curves, e.g. \cite{kent,things}.
There are two important differences. Firstly, it has a logarithmically increasing density profile
in the inner region, and secondly, it is constrained as 
$r_0$ is not a free parameter but set by the baryonic disk scale length. 
The logarithmically increasing inner density profile 
is expected to be observationally (virtually) indistinguishable 
from a truly flat profile, while the scaling of the core radius, $r_0 \sim r_D$,
is a noted feature derived from observations \cite{DS}.

Observe that the existence of a dark matter core, with a core radius $r_0 \sim r_D$, has a clear
geometrical origin in this dynamics.
The halo evolves towards a steady state configuration, which is strongly influenced
by the distribution of supernovae, as these represent the primary source of halo heating.
This heat source is cored given
the exponential distribution of the Freeman disk and the associated scale length, $r_D$.

\begin{figure}[t]
  \begin{minipage}[b]{0.5\linewidth}
    \centering
    \includegraphics[width=0.7\linewidth,angle=270]{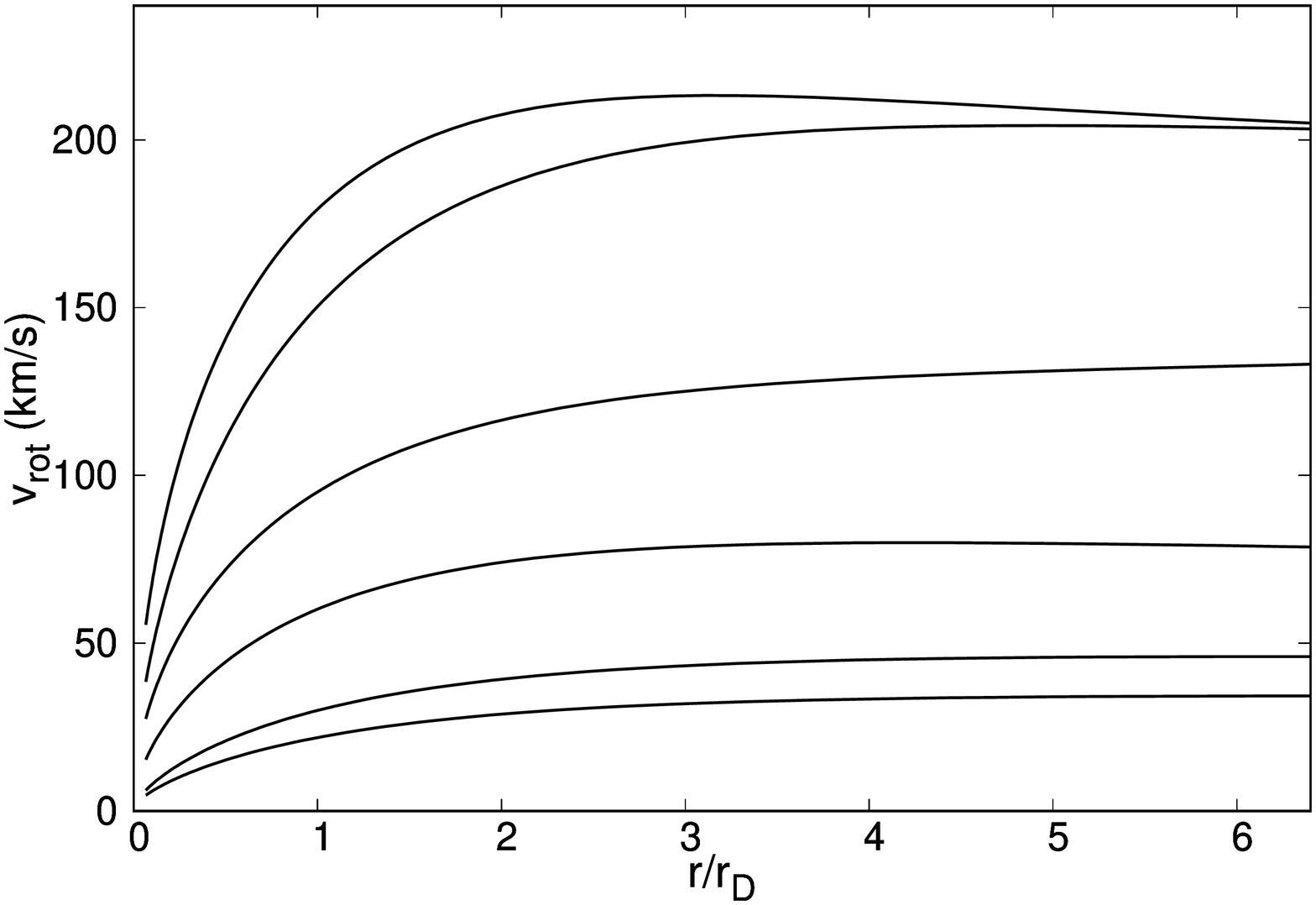}
     (a)
    \vspace{4ex}
  \end{minipage}
  \begin{minipage}[b]{0.5\linewidth}
    \centering
    \includegraphics[width=0.7\linewidth,angle=270]{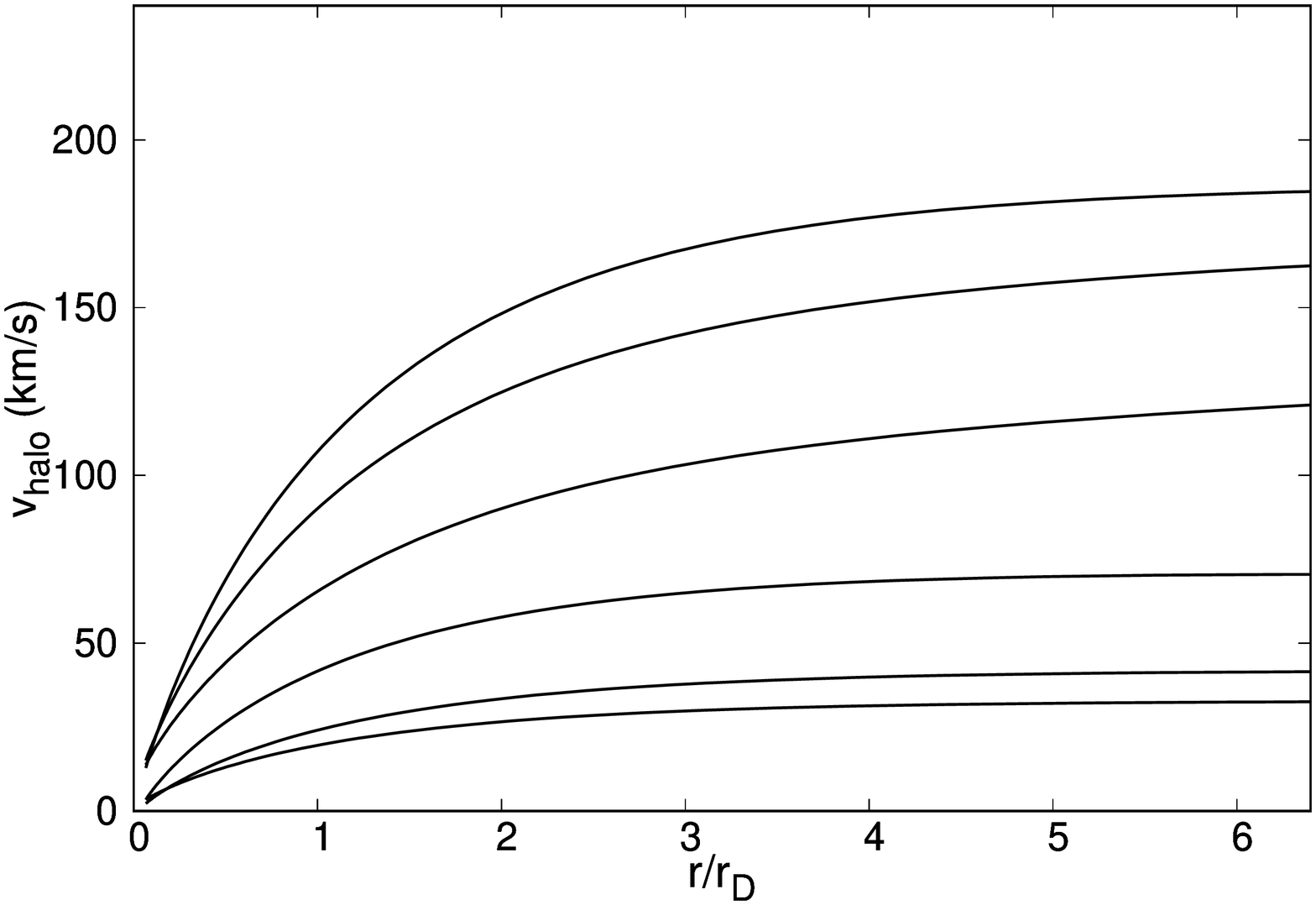}
    (b)
    \vspace{4ex}
  \end{minipage}
\vskip -1.0cm
\caption{
\small 
(a) the rotation curves (halo + baryons)
derived from the computed steady state solutions. 
The baryon mass ranges from
$m_{baryons} = 10^8 \ m_\odot$ (bottom curve) to $m_{baryons} = 10^{11} \ m_\odot$ (top curve).
See Table 2 for other baryonic parameters chosen. (b)
the corresponding halo rotation curves (halo contribution only).
}
\end{figure}

Consider now the rotation curves. 
The rotational velocity follows directly from Newton's law: 
\begin{eqnarray}
\frac{v_{rot}^2}{r} = \frac{G_N}{r^2} \int_0^r  [\rho(r') + \rho_{baryons}(r')] 4\pi r'^2 dr' \ . 
\end{eqnarray}
We are also interested in the dark halo contribution to the rotational velocity, for which we 
use the notation, $v_{halo}$: 
\begin{eqnarray}
\frac{v^2_{halo}}{r} = \frac{G_N}{r^2} \int_0^r  \rho(r') \  4\pi r'^2 dr' \ . 
\label{urg}
\end{eqnarray}
In Figure 5 we plot the rotational velocity and halo rotational velocity
derived from the steady state solutions found.

The rotation curves show an approximate linear rise in the inner region, turning over
to a roughly flat asymptotic profile, with the transition radius occurring at $r \sim r_D$
in each case.
As already mentioned, these properties can be understood from
simple geometrical considerations as the halo mass density is closely aligned with the
distribution of supernova sources.
Also, recall that these three properties are all well discussed features of measured rotation
curves.

\begin{figure}[t]
  \begin{minipage}[b]{0.5\linewidth}
    \centering
    \includegraphics[width=0.7\linewidth,angle=270]{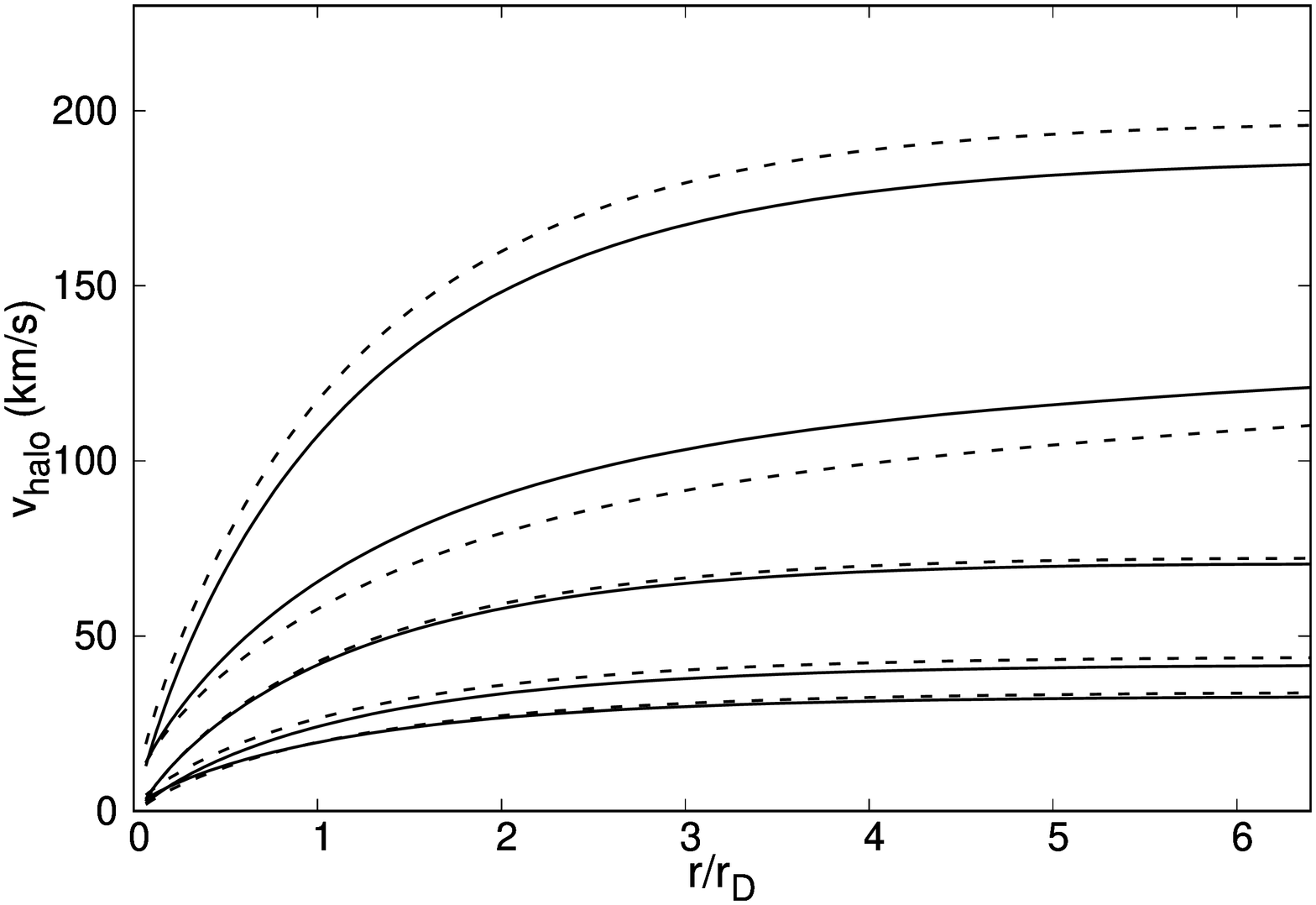}
     (a)
    \vspace{4ex}
  \end{minipage}
  \begin{minipage}[b]{0.5\linewidth}
    \centering
    \includegraphics[width=0.7\linewidth,angle=270]{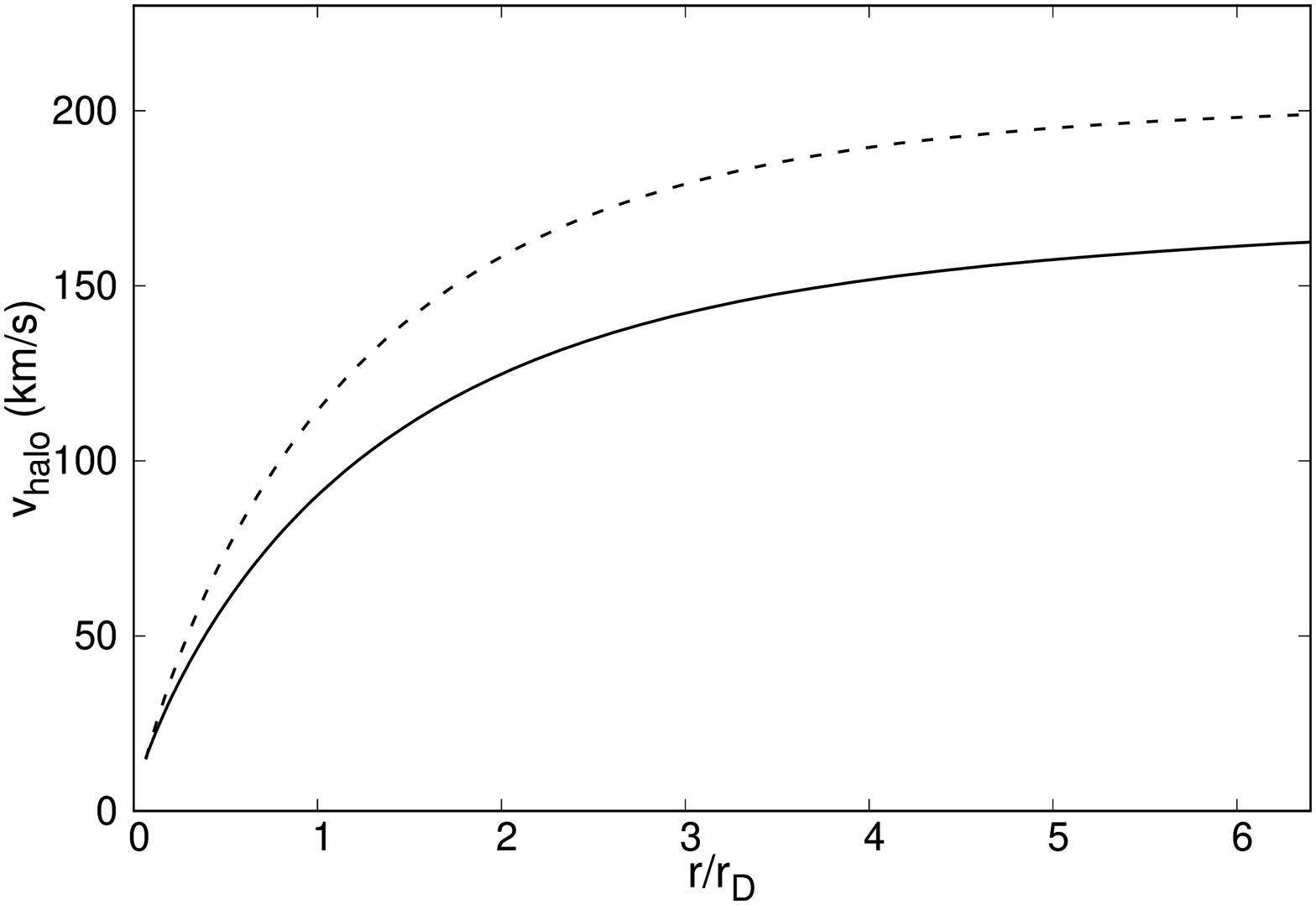}
    (b)
    \vspace{4ex}
  \end{minipage}
\vskip -1.0cm
\caption{
\small 
Halo rotation curves 
derived
from the steady state solutions for the galaxy set of  
Table 2 (solid lines),
and with a factor of $\times 2.5$ increase in the baryonic disk scale length (dashed lines).
(a) gives results for the five smallest galaxies, with
$m_{baryons} = 10^8 \ m_\odot$ (bottom curve) to $m_{baryons} = 10^{10.5} \ m_\odot$ (top curve),
while (b) is for $m_{baryons} = 10^{11} \ m_\odot$.
}
\end{figure}

We have also looked at the effects of varying some of the 
other baryonic parameters.
The baryonic parameters given in Table 2 are typical of high surface brightness galaxies.
In addition, there exist numerous low
surface brightness galaxies within which the stars and gas 
have a more extended spatial distribution.
To explore these kinds of galaxies, we have computed the
steady state solutions for the six galaxies of Table 2, but with disk
scale length increased
by a factor of $\times 2.5$ (i.e. $r_D \to 2.5r_D$, $r_D^{gas} \to 2.5r_D^{gas}$,
with $m_{baryons}, \ M_{FUV}, \ f_s$ unchanged). 
In Figure 6 we give the halo rotation curves derived from the steady state solutions
for the original (Table 2) parameter set, as well as the $r_D \to 2.5r_D$ variation.
As the figure shows, there is only a very minor effect on the halo contribution to the rotation curve
for the three smallest galaxies studied, while there are noticeable, although still modest, 
effects for the larger galaxies.
That is, the halo rotation velocity is, at least approximately, a function of the 
dimensionless variable $r/r_D$ (rather than
$r$ and $r_D$ separately).
This feature is consistent with observations, with 
the galaxies NGC2403 and UGC128  providing 
a well studied  illustration \cite{hsb}
(see also
\cite{stacy} for a recent discussion).

The near  
invariance of the
halo rotational velocity with respect to the transformation:
$r \to \Lambda r, \ r_D \to \Lambda r_D$, 
is not unexpected.
Recall the simple analytic argument that motivated the $\lambda$-density profile [Eq.(\ref{r1})]: 
for an optically thin isothermal halo,
${\cal H}(r) = {\cal C}(r)$ implies  a halo density proportional to the flux of SN 
sourced dark photons, i.e. $n(r) \propto F(r)$.
Since a photon flux scales as $\sim 1/r^2$, this suggests that the halo density
will scale as:
$\rho (r) \to \rho (\Lambda r)/\Lambda^2$, when $r \to \Lambda r, \ r_D \to \Lambda r_D$.
It follows directly from Newton's law, Eq.(\ref{urg}), that such a scaling implies
a scale invariant rotational velocity: $v_{halo}(r) \to v_{halo}(\Lambda r)$.
In fact,  the scaling:
$\rho (r) \to \rho (\Lambda r)/\Lambda^2$, when $r \to \Lambda r, \ r_D \to \Lambda r_D$,
is an exact property of the $\lambda$-density profile
(for fixed $\lambda$), which arises even
when the assumption of spherical symmetry is dropped \cite{olga}.
Of course,
scale invariance 
can only be
approximate, rather than exact, due to the effects of halo reabsorption (finite optical
depth) and departures from isothermality.

\begin{figure}[t]
\begin{minipage}[b]{0.5\linewidth}
  \centering
  \includegraphics[width=0.7\linewidth,angle=270]{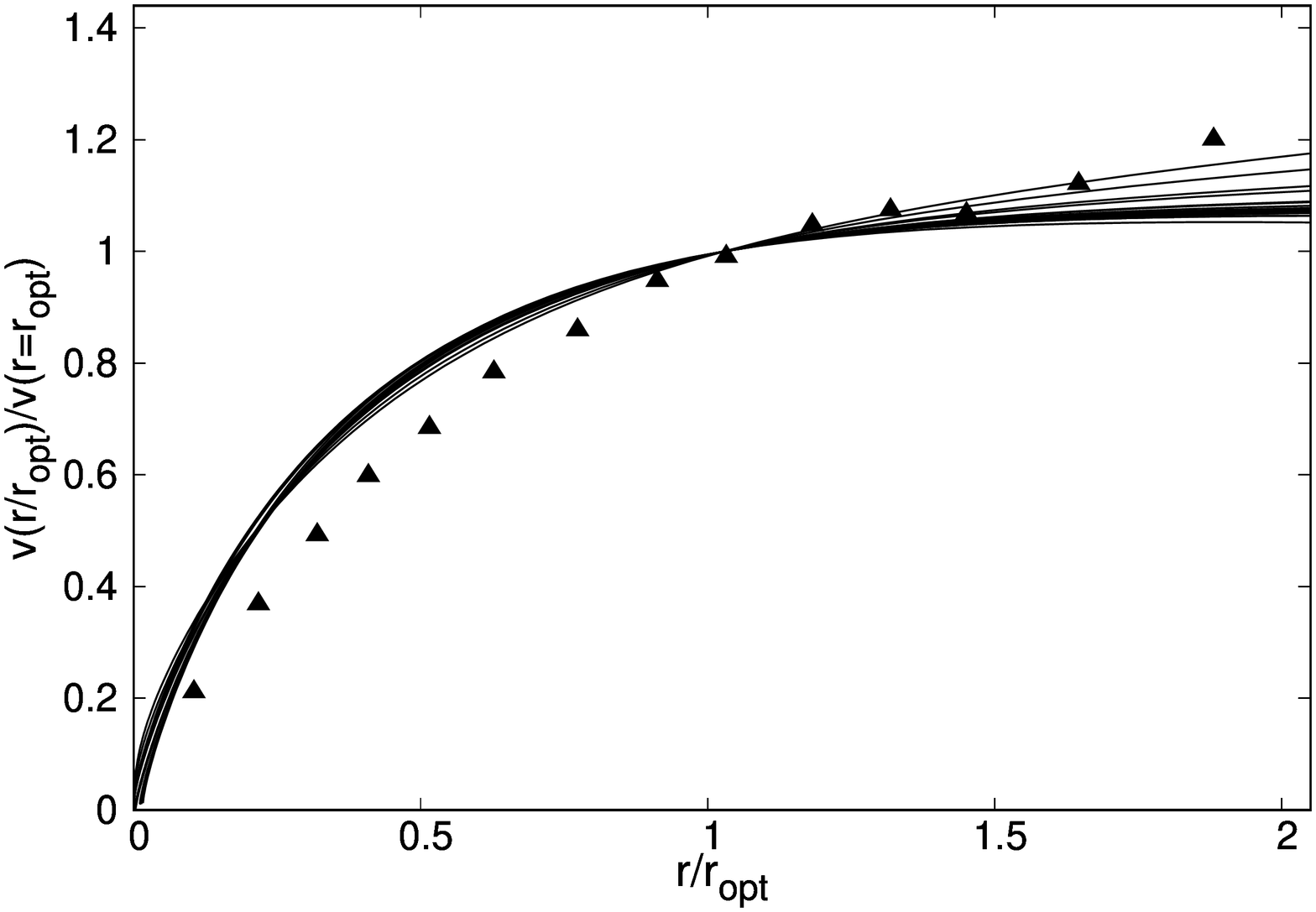}
   (a)
  \vspace{4ex}
\end{minipage}
\begin{minipage}[b]{0.5\linewidth}
  \centering
  \includegraphics[width=0.7\linewidth,angle=270]{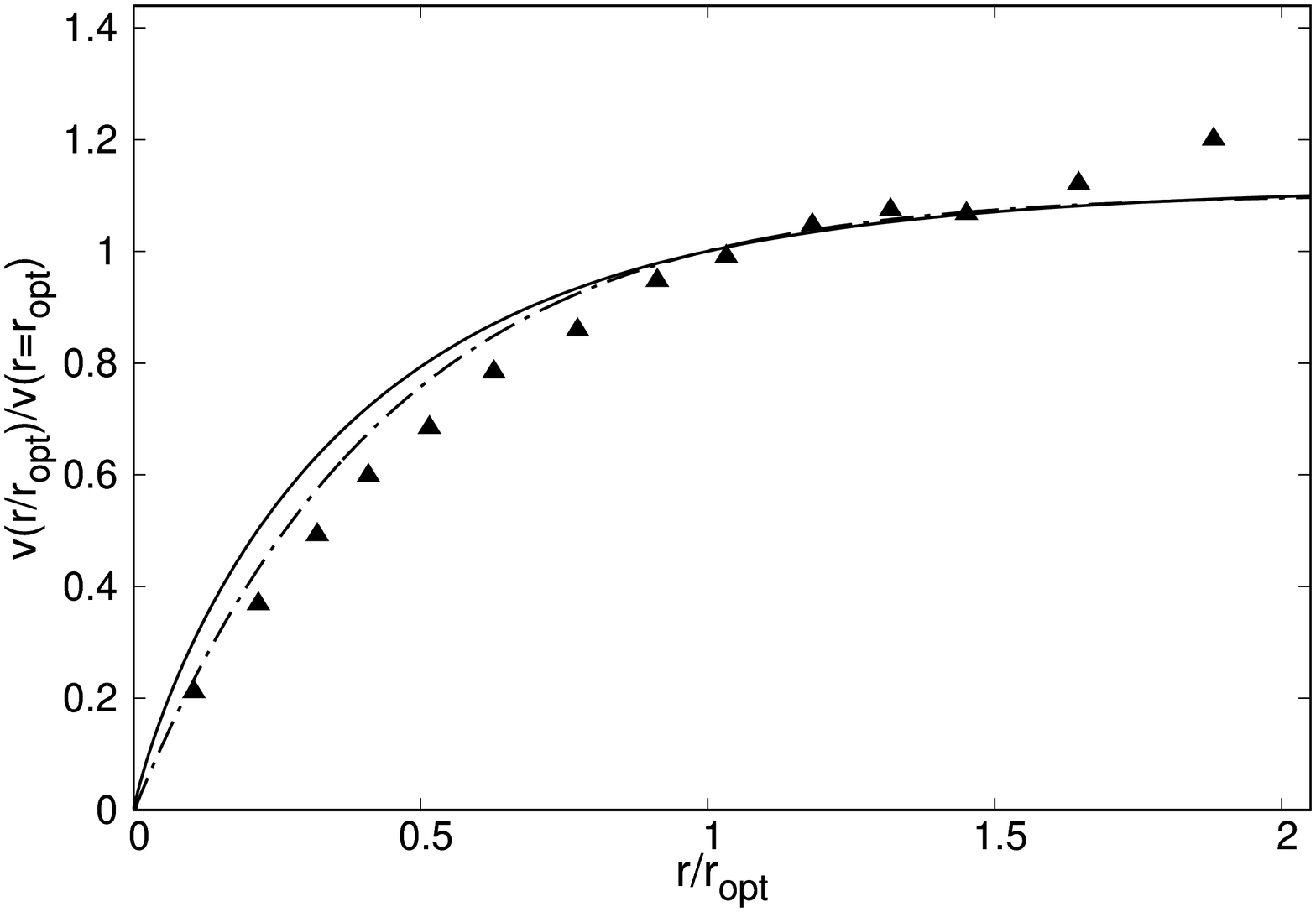}
  (b)
  \vspace{4ex}
\end{minipage}
\vskip -1.0cm
\caption{
\small
(a) Normalized halo rotational velocity, $v_{halo}(r)/v_{halo}(r_{opt})$, for all 18 modelled galaxies
calculated from the steady state solution.
(b) Comparison of the spherically symmetric $\lambda$-density profile [Eq.(\ref{r1})] 
(solid line)
with the corresponding profile for a disk geometry [Eq.(\ref{3z})] (dashed-dotted line).
The triangles in the figures are the synthetic rotation curve obtained from dwarf disk galaxies \cite{sal17}.
}
\end{figure}

Consider now the effect of varying the  luminosity while keeping the
other baryonic parameters fixed.\footnote{
The effect on the halo rotational velocity of varying 
the stellar mass fraction, $f_s$, with the other parameters kept 
fixed, was found to be 
negligible.}
Specifically, we have computed the steady state solutions  
for more luminous galaxies: $M_{FUV} \to M_{FUV} - 0.8$ 
(i.e. a factor of $10^{0.32} \approx 2.1$ increase in luminosity),
with the other baryonic parameters ($m_{baryons}$, $f_s$, $r_D$) 
unchanged from their canonical (Table 2) values. 
This set of baryonic parameters, together with the canonical parameter choice, and
those with $r_D \to 2.5r_D$, provide a total of 18 distinct galaxy parameters.
While changing the FUV luminosity will strongly influence the normalization
of the halo rotational velocity, let us first look at the effect (if any)
on the shape of the rotation curve.
Figure 7 shows the
derived
normalized halo rotational velocity, 
$v_{halo}(r)/v_{halo}(r_{opt})$, for all 18 modelled galaxies.
[Here $r_{opt} \simeq 3.2r_D$ is the optical radius.]
Evidently,
the halo rotation curves, 
that follow from the solution of the steady state conditions, Eq.(\ref{SSX}),
have a near universal profile, a notable feature consistent with observations, e.g.
\cite{stacy,sal17}. 
This result was anticipated given the major influence of  supernova sourced heating on the
halo density 
distribution \cite{foot16,olga}. The shape reflects the geometry
of the heating sources.

The steady state solutions discussed here correspond  to
galaxies simplified
with spherical symmetry.
Indeed, the supernova source distribution has been artificially modified
to make it spherically symmetric. 
In actuality, the spatial distribution of these 
sources is far from spherical;
a disk geometry would be much closer to realistic.
Nevertheless, we anticipate that
only modest changes would arise if
the system of equations were  solved without the spherically symmetric simplification.
In fact, the azimuthally symmetric disk analogue 
of the spherically symmetric $\lambda$-density profile 
[Eq.(\ref{r1})] is \cite{foot16,olga}:
\begin{eqnarray}
\rho(r,\theta) =
\lambda \int
d\widetilde{\phi} \int d\widetilde{r} \ \widetilde{r}
\ \frac{\Sigma_{SN}
({\widetilde{r}})} {4\pi[r^2 + {\widetilde{r}}^2 -
2r\widetilde{r}  \sin\theta \cos
\widetilde{\phi}]}
\label{3z}
\end{eqnarray}
where $\Sigma_{SN}$ is the supernova distribution
(averaged over a suitable timescale) in the disk.
The normalized halo rotational velocity  corresponding to this density,
shown  in Figure 7b, closely resembles
its spherically symmetric counterpart, and in fact agrees slightly better  with
the observations. 
While this appears to be a strong indication,
and indeed is very encouraging,
it still requires
verification that
Eq.(\ref{3z}) 
actually does approximate the steady state solution for galaxy systems with disk geometry.

\begin{figure}[t]
  \begin{minipage}[b]{0.5\linewidth}
    \centering
    \includegraphics[width=0.7\linewidth,angle=270]{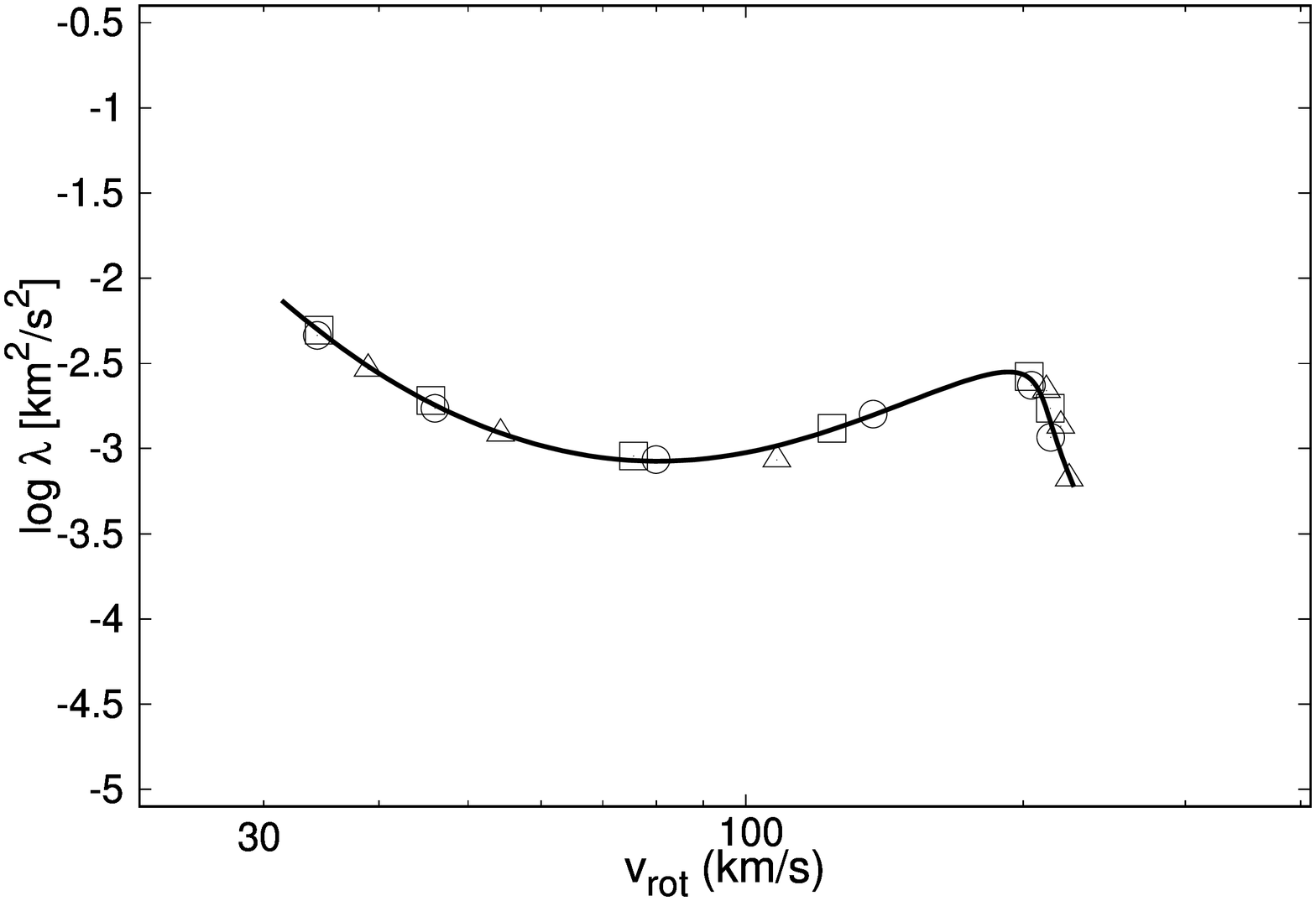}
     (a)
    \vspace{4ex}
  \end{minipage}
  \begin{minipage}[b]{0.5\linewidth}
    \centering
    \includegraphics[width=0.7\linewidth,angle=270]{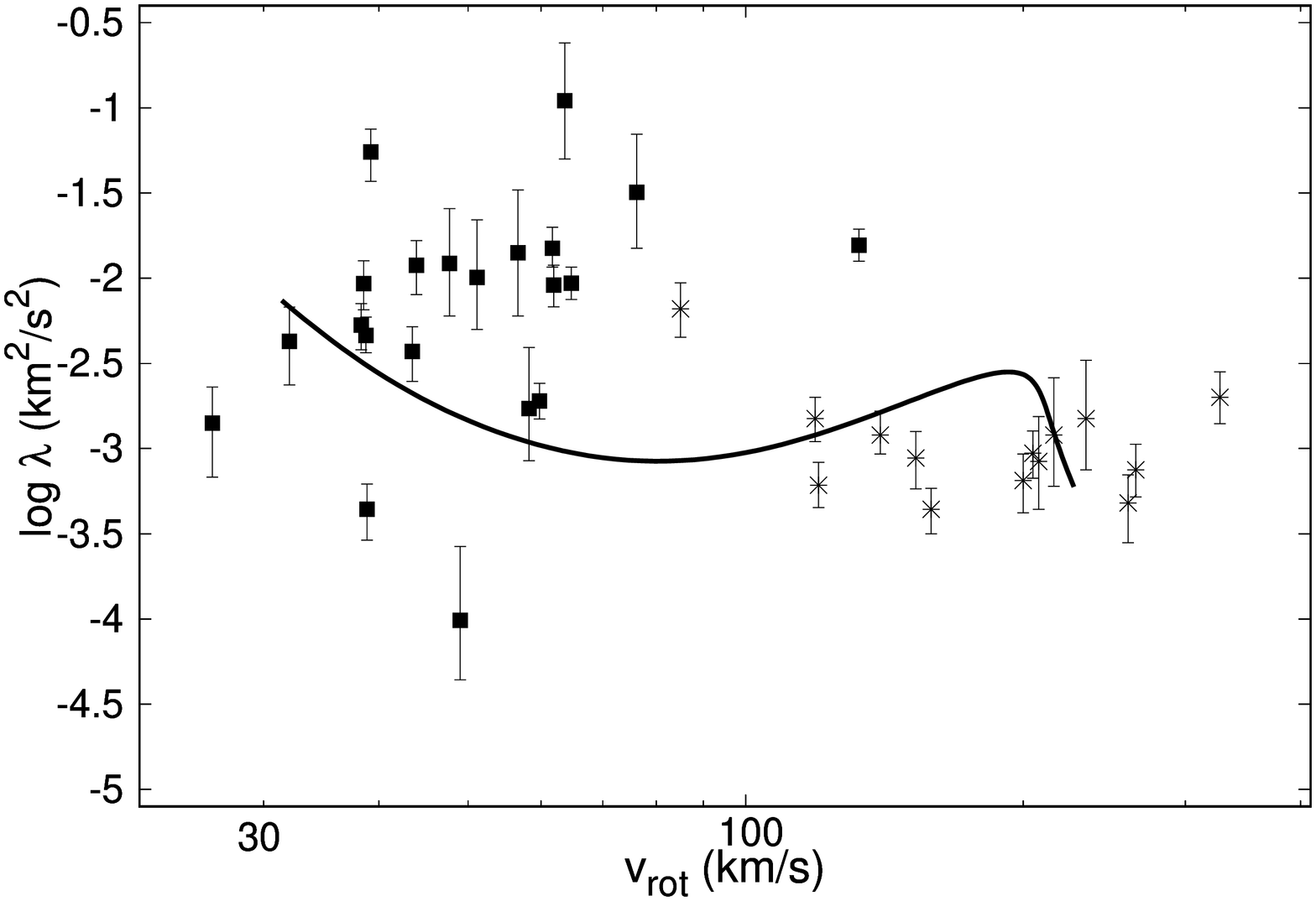}
    (b)
    \vspace{4ex}
  \end{minipage}
\vskip -1.0cm
\caption{
\small
The quantity $\widetilde{\lambda}$ [Eq.(\ref{10x})] 
versus $v^{max}_{rot}$ computed from the steady state solutions found. 
(a) gives the results for the 18 modelled galaxies 
along with an extrapolation
(circles for the  six canonical baryonic parameters
of Table 2, squares for the $r_D \to 2.5r_D$ parameter variation, and triangles for the
$M_{FUV} \to M_{FUV} - 0.8$  variation).
(b)  the extrapolated results together with 
the $\widetilde{\lambda}$ values from THINGS spirals (stars) \cite{things} 
and LITTLE THINGS dwarfs (squares) \cite{littlethings}.
}
\end{figure}

The next item of interest is
the scaling of the normalization of the halo velocity.
From Eq.(\ref{bnb}) we expect, at leading order, 
an asymptotic halo velocity of $v_{halo}^{asym} = 
\sqrt{G_N \lambda \kappa}$, 
i.e. $\lambda \propto [v_{halo}^{asym}]^2/R_{SN}$.
To make contact with observable quantities, we again make use of the expected
$R_{SN} \propto L_{FUV} \propto 10^{-0.4 M_{FUV}}$ scaling.
It is convenient then, to introduce the quantity $\widetilde{\lambda}$: 
\begin{eqnarray}
\widetilde{\lambda} \ \equiv \ \frac{[v^{asym}_{halo}]^2}{10^{-0.4 M_{FUV}}}
\label{10x}
\ .
\end{eqnarray}
We have evaluated $\widetilde{\lambda}$ from the computed steady state
solutions for all 18 galaxy parameters examined [taking $v^{asym}_{halo} = v_{halo}(r=6.4r_D)$].
The result of 
this exercise is shown in Figure 8, where we plot the  obtained $\widetilde{\lambda}$ values
versus the maximum of the rotational velocity, $v_{rot}^{max}$. Also shown in the figure
are the values of $\widetilde{\lambda}$  for THINGS spirals \cite{things} and 
LITTLE THINGS dwarfs \cite{littlethings}.\footnote{
The raw FUV absolute magnitude values were appropriated from the 
updated nearby galaxy catalogue \cite{table}, and 
corrected for internal and foreground extinction following \cite{uv3,uv2}.
}

\begin{figure}[t]
  \begin{minipage}[b]{0.5\linewidth}
    \centering
    \includegraphics[width=0.7\linewidth,angle=270]{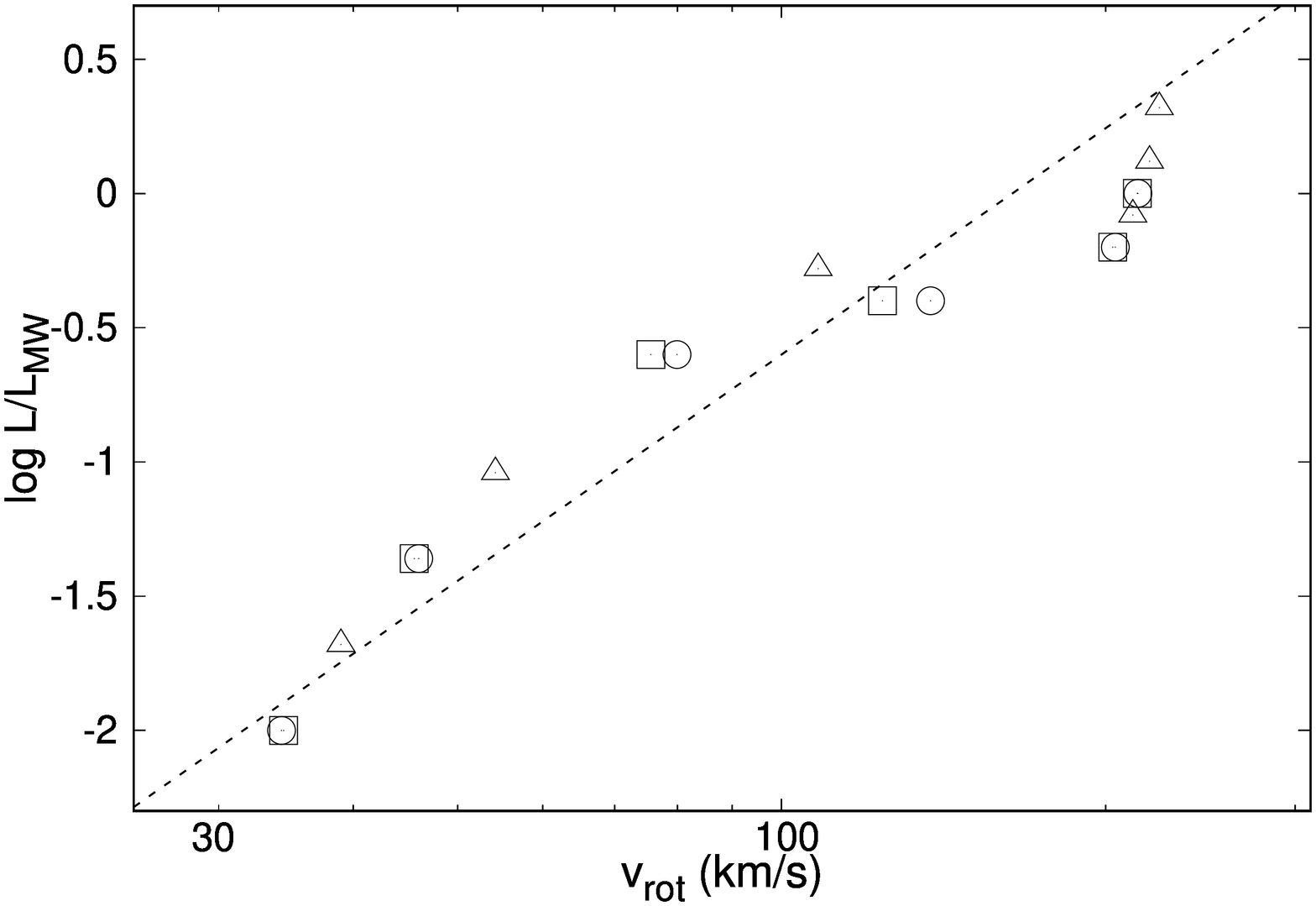}
     (a)
    \vspace{4ex}
  \end{minipage}
  \begin{minipage}[b]{0.5\linewidth}
    \centering
    \includegraphics[width=0.7\linewidth,angle=270]{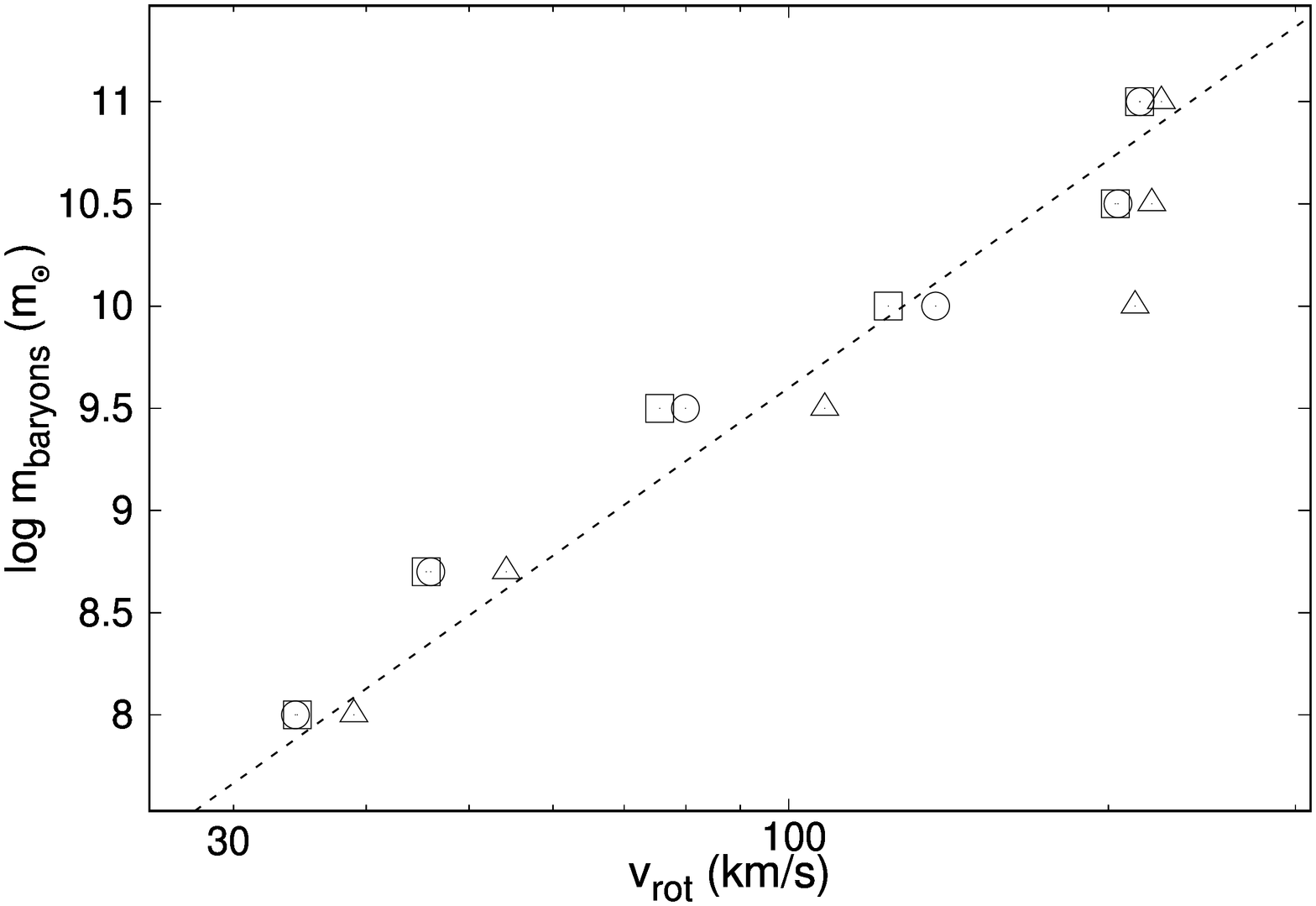}
    (b)
    \vspace{4ex}
  \end{minipage}
\vskip -1.0cm
\caption{
\small
(a)  $L_{FUV}/L^{MW}_{FUV}$ versus $v^{max}_{rot}$
and (b) $m_{baryons}$ versus $v^{max}_{rot}$, for each of the steady state solutions found. 
Circles  are the results for the canonical galaxy set of
Table 2, squares for the $r_D \to 2.5r_D$ parameter variation, and triangles for the
$M_{FUV} \to M_{FUV} - 0.8$  variation.
The dashed lines are the power laws:  (a) $L_{FUV} \propto [v_{rot}^{max}]^{2.8}$ and
(b) $m_{baryons} \propto [v_{rot}^{max}]^{3.7}$.
}
\end{figure}

Figure 8 indicates that, for the model galaxies studied, $\widetilde{\lambda}$ is
not exactly constant but has some variation 
with respect to $v^{max}_{rot}$.\footnote{The sharp downturn at $v_{rot}^{max} \sim
200$ km/s is a threshold effect. The plasma  is approaching the state of full ionization. 
Of course, a change in fundamental parameters, e.g. decreasing $m_{p_d}$ (which lowers the halo
temperature), can move this threshold to a higher $v_{rot}^{max}$ value.}
The overall normalization  appears to be in the right ballpark to be consistent
with THINGS spirals and LITTLE THINGS dwarfs, although the 
$\widetilde{\lambda}$  values of the dwarfs show 
significant scatter. The rotation curve shapes of many of the LITTLE THINGS
dwarfs are also quite irregular.
This may be an indication that many of these small galaxies are 
not currently in a steady state configuration. In fact,  
some of these  galaxies are known to  have unstable star formation (recent) histories. They
are starburst galaxies, undergoing large scale oscillations in star formation rate,
with a period of order $\sim 100$ Myr, e.g.\cite{herr,uv3}.\footnote{In this picture,
the oscillations in the rate of star 
formation  might be
strongly influenced by nontrivial dissipative halo dynamics. Large scale 
radial density oscillations of 
the plasma halo will induce oscillations in the star formation rate,  due to
the expansion and compression of the baryonic gas
under the oscillating gravitational field strength.
Such phenomena appear extremely interesting, but complex, requiring 
solution of the time dependent fluid  
equations.
}

The results for the halo rotational velocity normalization (Figure 8)  
together with our earlier results for the shape (Figure 7) can be summarized:
The halo rotation curve
that follows  from the steady state conditions
has a characteristic functional form that depends (approximately) only on $r/R_D$, $M_{FUV}$.
This appears to be consistent with observations
- with the existence of such a
characteristic functional form discussed for many years,
e.g. \cite{PS91,PPS}. 
It
has also been argued, though, that the
normalization of the rotational velocity might depend more 
closely on $m_{baryons}$ (rather than luminosity), with 
$m_{baryons} \propto [v_{rot}^{max}]^{\beta}$, $\beta \approx 4$ 
(baryonic Tully-Fisher relation \cite{btf}).

In Figure 9a [9b] we plot $L_{FUV}$ [$m_{baryons}$] versus $v_{rot}^{max}$ for
the modelled galaxies.
Evidently, the results are broadly compatible 
with the empirical scaling relations, but with some
scatter. Clearly though, the underlying relation is a 
predicted scaling of $\widetilde{\lambda}$ (Figure 8),
for which there is negligible scatter. (We have checked this further by considering a wider
variation of baryonic parameters than those displayed.)
That is, our analysis indicates that a  
$\widetilde{\lambda}$ versus $v_{rot}^{max}$ correspondence
underpins the empirical  Tully-Fisher
relations, at least in the kind of dissipative model discussed here. 
Naturally, this connection could be examined more closely in future studies, noting
that the weak variation of $\widetilde{\lambda}$ with $v_{rot}^{max}$  will surely
have some dependence on the fundamental parameters defining
the dissipative dark matter model.

Other dark matter galactic scaling relations, e.g. 
the radial acceleration relation \cite{MC,Lelli},
can be viewed as a consequence of the predicted universal profile (Figure 7)
with correct normalization (Figure 8), see discussion in e.g. \cite{olga}.
Further exploration of such relations is therefore not essential.

\begin{figure}[t]
  \begin{minipage}[b]{0.5\linewidth}
    \centering
    \includegraphics[width=0.7\linewidth,angle=270]{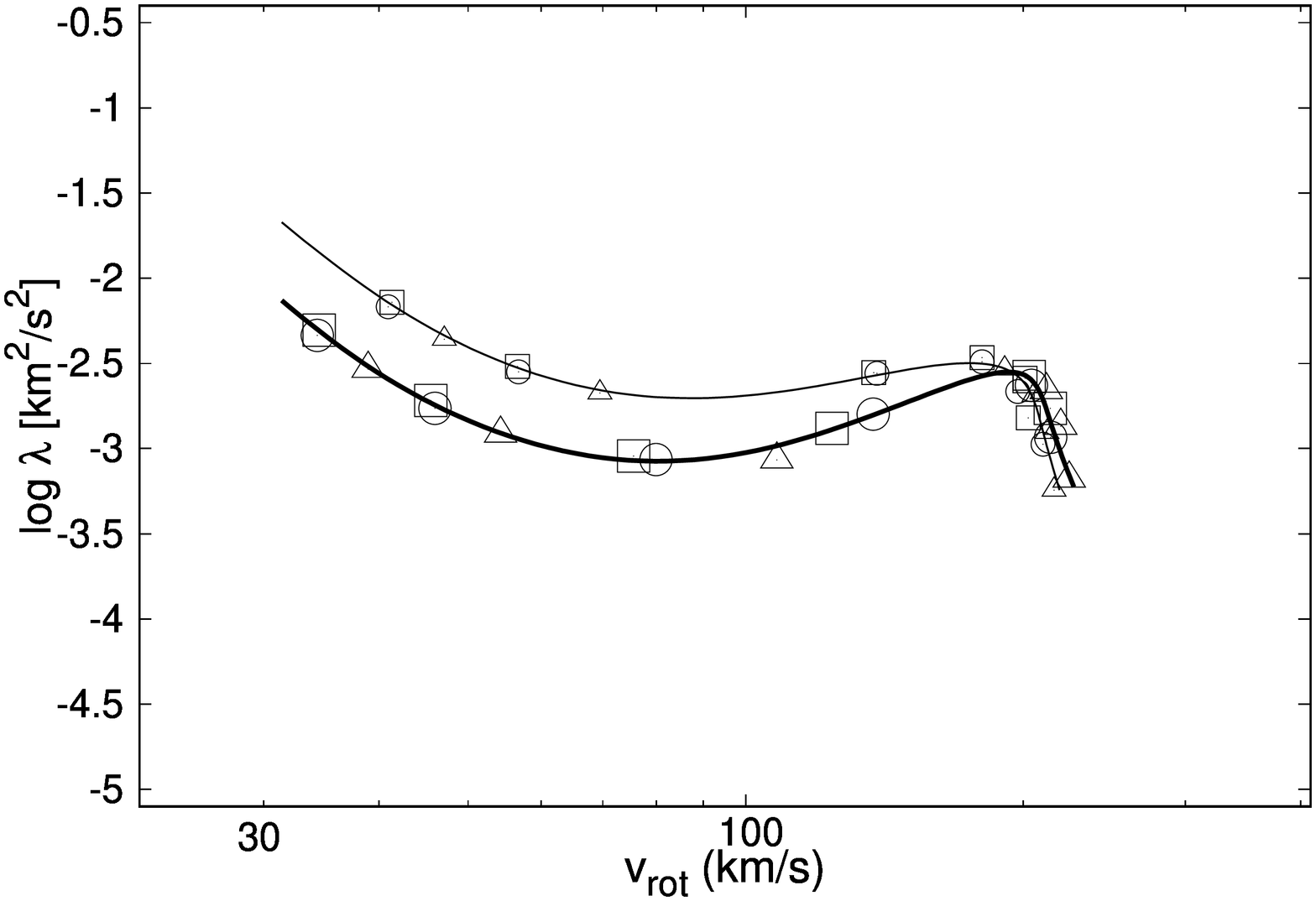}
     (a)
    \vspace{4ex}
  \end{minipage}
  \begin{minipage}[b]{0.5\linewidth}
    \centering
    \includegraphics[width=0.7\linewidth,angle=270]{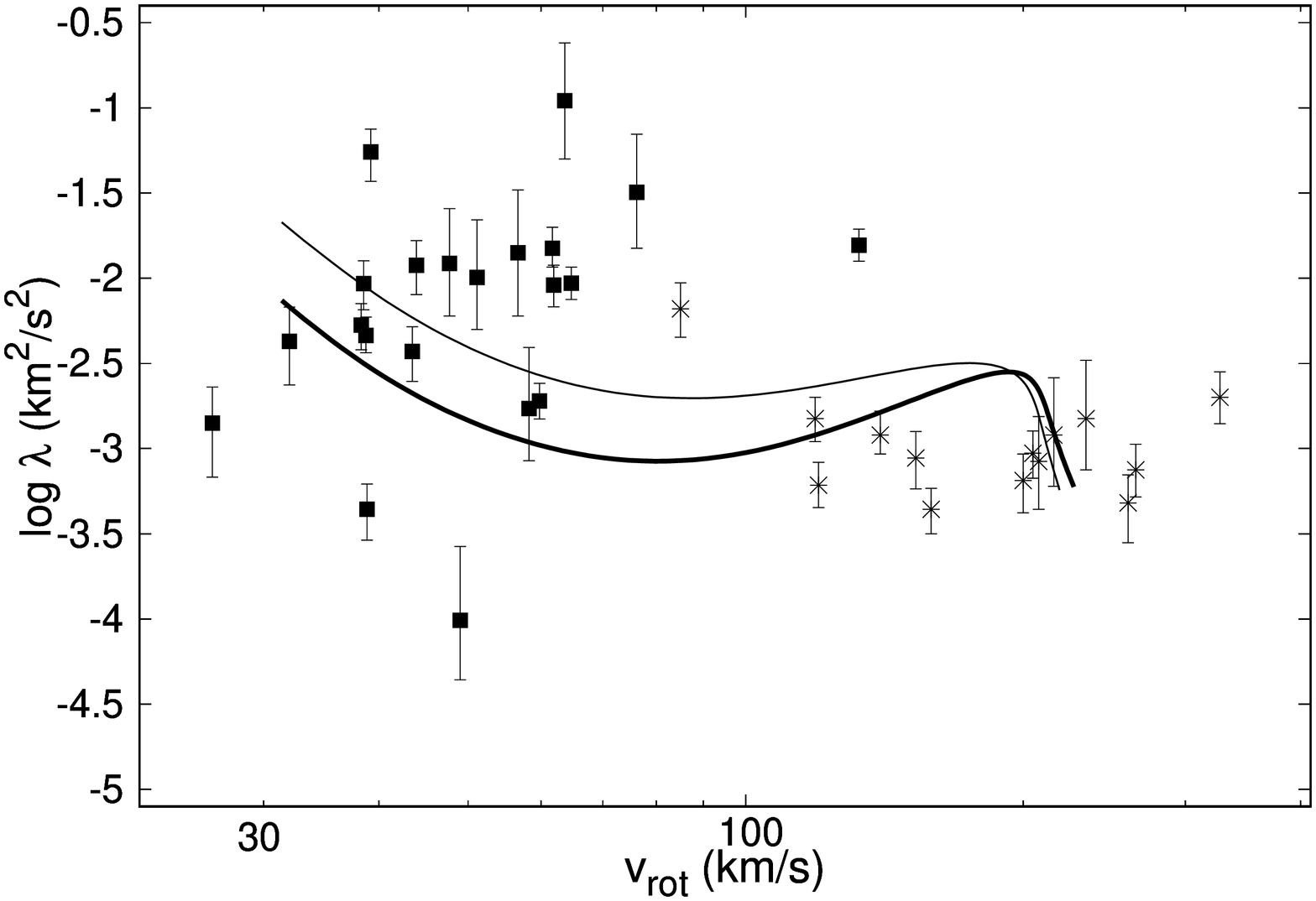}
    (b)
    \vspace{4ex}
  \end{minipage}
\vskip -1.0cm
\caption{
\small
(a) The effect of varying the effective temperature parameter, $T_{eff}$ [Eq.(\ref{isis})],
on  $\widetilde{\lambda}$, Eq.(\ref{10x}). 
Notation as in Figure 8a but with small (large) 
symbols for $T_{eff} = 10$ keV ($T_{eff} = 25$ keV). 
(b) the extrapolated results
for  $T_{eff} = 10$ keV (thin solid line) and  $T_{eff} = 25$ keV (thick solid line).
Also shown are
the $\widetilde{\lambda}$ values from THINGS spirals (stars) \cite{things} 
and LITTLE THINGS dwarfs (squares) \cite{littlethings}.
}
\end{figure}

Finally, to complete this analysis of the particular dissipative model parameters chosen,
we briefly examine the effects of varying the uncertain effective supernova temperature
parameter, $T_{eff}$. [Recall, we have modelled the dark photon frequency spectrum
by a thermal distribution with this effective temperature, Eq.(\ref{isis}).]
Changing the $T_{eff}$ parameter 
will modify the level of halo heating
and, assuming a steady state solution still exists, will potentially influence 
the normalization of the halo velocity.
However, the shape of the normalized halo rotational velocity remains (approximately)
unchanged, and simply reflects the geometry of the SN heat source distribution (as discussed).
Figure 10 compares 
the
$\widetilde{\lambda}$ [Eq.(\ref{10x})] 
values obtained with 
two choices for the
effective temperature parameter, $T_{eff} = 10$ keV and $T_{eff} = 25$ keV,
for the same 18
baryonic galaxy parameters already discussed.

\section{Conclusion}

We have considered dark matter featuring particle properties that closely resemble 
familiar baryonic matter.
Mirror dark matter, which arises from an isomorphic hidden sector, is a specific and theoretically constrained
scenario. 
Other possibilities include models with more generic 
hidden sectors that contain massless dark photons (unbroken $U(1)$ gauge interactions).
Such dark matter not only features dissipative cooling processes, but is assumed to have 
nontrivial heating sourced by ordinary supernovae (facilitated by the kinetic mixing interaction).

The dynamics of dark matter halos around rotationally supported galaxies, influenced
by cooling and heating processes, can be modelled by fluid equations. 
For a sufficiently isolated galaxy with
stable star formation rate,
the dissipative dark matter halos are expected 
to evolve to a steady state configuration which is in hydrostatic
equilibrium and 
where heating and cooling rates locally balance.
Here, we have endeavored  to take into 
account the major cooling and heating processes, and 
have numerically solved for the
steady state solution under the assumptions of spherical symmetry, 
negligible dark magnetic fields,
and that supernova sourced energy is transported to the halo
via dark radiation.
For the parameters considered, and assumptions made, 
we were unable to find a physically realistic solution for 
the theoretically constrained case of mirror dark matter halos. 
Halo cooling generally exceeds heating at realistic halo mass densities.

Naturally, there are a number of possible reasons for this 
deficiency, some of the assumptions made could be re-examined etc. 
It could also be that mirror dark matter is not the correct dark matter model. 
Nature might prefer some other kind of
dissipative dark matter, if dark matter is indeed dissipative.
To illustrate such a possibility,
we examined galaxy structure in the context
of more generic dissipative dark matter models.
One such model was looked at in some detail which featured steady state solutions with
realistic dark matter halos.
This analysis confirmed, to some extent, the insight gleaned from simplified analytical
considerations, e.g. \cite{sunny1,foot16}.
 
We conclude this work by summarizing 
some of the key results from this analysis of the steady state
solutions of the particular dissipative model studied. 
The rotation curves are characterized by three main features:
\begin{itemize}
\item Approximate scale invariance of the halo rotation velocity: $v_{halo}(r) \to v_{halo}(\Lambda r)$
under $r \to \Lambda r, \ r_D \to \Lambda r_D$ (Figure 6).

\item The shape of the normalized halo rotational velocity is close to universal (Figure 7).

\item The normalization of the halo velocity is characterized 
by a Tully-Fisher type relation (Figure 8).
\end{itemize}
Even though  this analysis has focused on a particular dissipative model, the above three
features are expected to hold over a significant region of parameter space
in generic dissipative models \cite{foot15,foot16,sunny1,olga}. 
Observations appear to be consistent with the above features, including the universal profile 
for the normalized halo rotational velocity
that was here derived in the steady state limit. 
That observations have these characteristic features has been frequently noted 
in the literature, and is often cited
as support for the notion of modified Newtonian dynamics (MOND) \cite{mond1,mond2,mond3}. 
The present study, though,  reinforces the idea that
(approximate) MONDian phenomenology can arise also in a dark matter setting.

\vskip 1.7cm
\noindent
{\large \bf Acknowledgments}

\vskip 0.2cm
\noindent
The author would like to acknowledge the hospitality of the IHEP in Beijing where
some of this work was undertaken.
This work was supported by the Australian Research Council.

\end{document}